\documentclass[12pt, journal, onecolumn]{IEEEtran} 
\pdfoutput=1
\usepackage{amsfonts}
\usepackage{amsmath} 
\usepackage{amssymb} 
\usepackage{amsthm,amscd}
\usepackage{bm} 
\usepackage{graphicx,epsfig}
\usepackage{eucal}      
\usepackage{cite}
\usepackage{enumerate}
\usepackage{verbatim}       
\usepackage{makeidx}        
\usepackage{url}        
\usepackage[tight,footnotesize]{subfigure}
\usepackage{algorithm}
\usepackage{algorithmicx}
\usepackage{algpseudocode}
\usepackage[margin=1 in]{geometry}
\usepackage{tabularx}
\allowdisplaybreaks[4]
\usepackage{hyperref}
\newtheorem{lem}{Lemma}

\def\E{\mathbb{E}}
\def\EE{\mathbb{E}^{!o}}
\def\P{\mathbb{P}}

\def\R{\mathbb{R}}

\def\L{\mathcal{L}}

\def\d{\mathrm{d}}

\def\I{\mathtt{I}}

\def\PP{\mathbb{P}^{!o}}

\def\EP{\mathbb{E}^{!o}}
\def\T{\theta}

{}
  
  \newtheorem{theorem}{Theorem}

\def\PP{{\mathbb{P}^{!o} }}
\def\EE{{\mathbb{E}^{!o} }}
\def\R{{\mathbb{R} }}
\def\E{{\mathbb{E} }}

\def\d{\mathrm{d}}

\makeatletter
\newcommand{\biggg}{\bBigg@{2}}
\newcommand{\Biggg}{\bBigg@{3}}
\newcommand{\Bigggg}{\bBigg@{4}}
\makeatother

\begin{document}

\title{MIMO Interference Alignment in Random Access Networks}
\author{Behrang~Nosrat-Makouei,
        		Radha~Krishna~Ganti,
                Jeffrey~G.~Andrews,
                and Robert~W.~Heath,~Jr.,
\thanks{Behrang~Nosrat-Makouei, Jeffrey~G.~Andrews, and Robert~W.~Heath,~Jr., are with the Department of Electrical and Computer Engineering, The University of Texas at Austin, Austin, TX 78712 USA (e-mail:
behrang.n.m@mail.utexas.edu; jandrews@ece.utexas.edu; rheath@ece.utexas.edu). 
Radha~Krishna~Ganti is with the Department of Electrical Engineering, Indian Institute of Technology Madras, Chennai, India (e-mail:
rganti@ee.iitm.ac.in).}
\thanks{This work was supported by the DARPA IT-MANET program, Grant W911NF-07-1-0028, and by the Army Research Labs, Grant W911NF-10-1-0420.
This work was presented in part at Asilomar Conf. on Signals, Syst. \& Comput., Pacific Grove, CA, Nov. 2011 \cite{BehrangAsilomar2011}.}}
\maketitle

\begin{abstract}
In this paper, we analyze a multiple-input multiple-output (MIMO) interference channel where nodes are randomly distributed on a plane as a spatial Poisson cluster point process. Each cluster uses interference alignment (IA) to suppress intra-cluster interference but unlike most work on IA, we do not neglect inter-cluster interference. We also connect the accuracy of channel state information to the distance between the nodes, i.e. the quality of CSI degrades with increasing distance. Accounting for the training and feedback overhead, we derive the transmission capacity of this MIMO IA ad hoc network and then compare it to open-loop (interference-blind) spatial multiplexing. Finally, we present exemplary system setups where spatial multiplexing outperforms IA due to the imperfect channel state information or the non-aligned inter-cluster interference.
\end{abstract}

\section{Introduction}
Interference alignment (IA) achieves the degrees of freedom in the $K$-user MIMO interference channel \cite{Gou2010}. IA confines the interference to a subspace at each receiver such that an interference-free subspace becomes available for the desired signal transmission. Except for \textit{blind} IA techniques \cite{Jafar2012,Gou2011}, which usually attain a lower multiplexing gain, IA requires cooperation between the transmitting/receiving nodes using global channel state information (CSI) \cite{Cadambe2008a} or some form of channel reciprocity. Although the impact of inaccurate CSI on the performance of IA has been studied before \cite{BehrangTSP2011}, existence of non-cooperating nodes casting non-aligned interference has been mostly ignored. 

\subsection{Background}
\indent In large networks, such as mobile ad hoc and cellular networks, IA will be used independently in separate clusters and the nearby nodes that are not coordinating with any one cluster will cause non-aligned interference at the receivers. There are three main reasons. 
\begin{enumerate}
\item Number of antennas at each node is a limiting factor. It is shown in \cite{Gou2010} that the number of nodes that can cooperate through MIMO IA is limited by the number of antennas at each node. 
\item Overhead practically limits the cluster size. The overhead of IA grows super linearly with the number of users\cite{Peters2012}, and hence it is likely that small groups of nodes will coordinate to perform IA.
\item More cooperation is not always better. Recently, \cite{Lozano2012} showed that because of inherent channel uncertainty, there is a moderate cluster size above which spectral efficiency at best saturates, and in many practical scenarios (e.g. when pilots are used for channel estimation), actually decreases if more nodes join the cluster to cooperate.
\end{enumerate}
In this case, single cluster analysis (e.g. DoF studies) does not capture the impact of interference from the other nodes in the network and can lead to unrealistic cooperation gains which would not be attainable if the \textit{inter-cluster} interference was accounted for \cite{Lozano2012,baccelli2009_Book_Vol2}. Most work on the performance of IA, however, is confined to single-cluster performance analysis (see \cite{BehrangTSP2011,Ayach2012,BehrangTWireless2011,gollakota2009interference} and references therein).

\indent When dealing with large networks, a relevant metric of the system performance is the transmission capacity \cite{Weber2005}, defined as the number of successful transmission per unit area, subject to a constraint on outage probability. The transmission capacity, in contrast to other network-wide system performance metrics such as transport capacity, generally leads to closed-form expressions or tight bounds providing insight into the network design parameters \cite{WeberTC2010}. To the best of our knowledge, little prior work on the network-wide performance of MIMO IA systems exists. In \cite{Tresch2011}, the spatial distribution of nodes is taken into account for deriving the point-to-point outage probability but the accuracy of the acquired CSI is ignored. By assuming perfect CSI, however, as discussed in \cite[Section V]{BehrangTSP2011}, the authors in \cite{Tresch2011} effectively favor IA over other transmission techniques which either do not require CSI at the transmitters or are less sensitive to CSI imperfections. Therefore, the goal of this paper is take into account both the node distribution and the CSI uncertainty to better understand the performance of MIMO IA in large decentralized networks.

\subsection{Contributions}
\indent  In this paper, we find the transmission capacity of a large ad hoc network where nodes are partitioned into separate clusters each cooperating through IA. We assume a four-stage transmission protocol. In the first stage, with a finite length  \textit{training period}, imperfect CSI for the cross links is obtained through MMSE channel estimation.
In the second stage, the estimated CSI is fed back to the other nodes in the cluster during the \textit{feedback period}. In the third stage, the IA transmit/receive filters are computed.
In the last stage, using the rest of the finite-length channel block, the nodes communicate data using a cluster-wise slotted Aloha-like channel access protocol where at random, all nodes in a cluster either transmit simultaneously or turn off their transmission. MIMO IA, as discussed in this paper, requires synchronization and coordination in each cluster and therefore a coordinated decision to transmit or not is reasonable. 

\indent Toward deriving the transmission capacity of this network, we first derive the exact point-to-point outage probability at a typical receiver. Then, assuming fixed feedback overhead, we solve for the optimum training period locally maximizing each cluster's goodput. Next, we derive the transmission capacity and compare it to a network with the same topology where only a single transmit/receiver pair in each cluster utilizes spatial multiplexing at each time instant.

\indent Our results indicate that the transmission technique of choice is a function of the node density, the mobility of the nodes, the transmit power, and the characteristics of the underlying communications medium. 
For example, in dense networks with high transmit power, spatial multiplexing (SM) over an orthogonal channel access strategy such as time division multiple access (TDMA) can outperform IA due to lower inter-cluster interference.
Also, the signal-to-noise-ratio (SNR) switching point between IA and TDMA+SM decreases with increasing density and mobility. Our initial work in \cite{BehrangAsilomar2011} only deals with point-to-point outage probability. This paper elaborates on the claims of \cite{BehrangAsilomar2011}, solves for the optimum training period, finds the corresponding transmission capacity of the network, provides easier to compute bounds in several important cases, and presents new simulation results. 

\subsection{Organization and Notation}
\indent The remainder of the paper is organized as follows. In Section \ref{sec:system_model} we present the system model. In Section \ref{sec:IntraCluster_IA} we analyze the performance of intra-cluster MIMO IA through quantifying point-to-point outage probability and the transmission capacity. In Section \ref{sec:SM} we derive parallel performance metrics for the same network utilizing spatial multiplexing. We present numerical results in Section \ref{sec:NumericalResults} followed by concluding remarks in Section \ref{sec:Conclusion}. Also, Tables \ref{table:Notation} and \ref{table:symbols} present the notation used and the important symbols defined in this manuscript.

\section{System Model} \label{sec:system_model}
\begin{figure}[!b]
\centering
\includegraphics[width=4.3in]{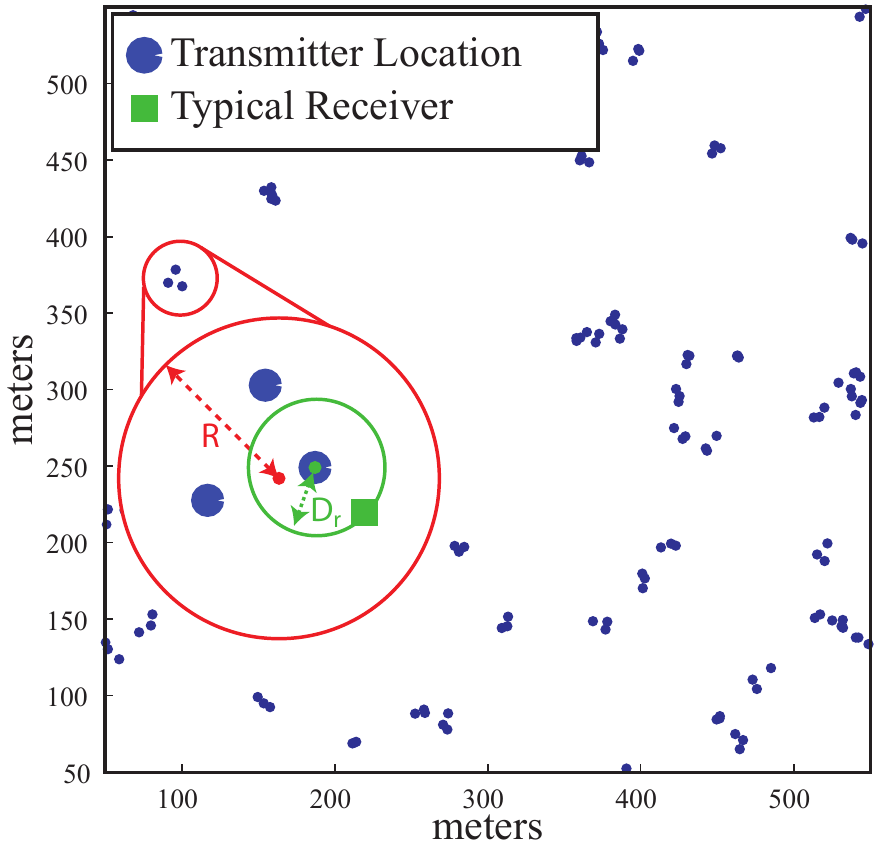}
\caption{An instance of the transmitter's distribution when the number of transmitters per cluster is $K=3$.} 
\label{fig:systemModel}
\end{figure}

\indent The spatial locations of the potential transmitters, $\boldsymbol{\Phi}$, are modeled as a planar Neyman-Scott cluster point process \cite{Stoyan2008}.  In this process, the cluster centers are modeled by a parent homogeneous Poisson point process (PPP) $\boldsymbol{\Phi}_{\rm{p}}$  of density $\tilde{\lambda}_{\rm{p}}$. Each parent point $x \in \boldsymbol{\Phi}_{\rm{p}}$ forms the center of a cluster around which $K$ daughter points are uniformly distributed in a circle of radius $R$. The resulting process\footnote{The parent points $\boldsymbol{\Phi}_{\rm{p}}$ will not be a part of the final point process.} is a stationary point process of density $K\tilde{\lambda}_{\rm{p}}$. We assume clusters randomly access the channel with probability $P_{\rm{A}}$ effectively reducing the density of this PPP to $\lambda_{\rm{p}}\!=\!P_{\rm{A}}\tilde{\lambda}_{\rm{p}}$. The receiver of a transmitter at $x$ is denoted by $\hat{x}$ and is assumed to be randomly located at distance $D_{\rm{r}}$ from its transmitter forming an $N\times N$ MIMO link. The receivers are  not part of the point process $\boldsymbol{\Phi}$. An instance of the nodes' location is shown in Fig. \ref{fig:systemModel}.

\indent In this paper, a typical transmitter (a transmitter chosen at random) is considered and its performance is analyzed.
This transmitter is typical in the sense that the performance of IA in this node is a representative of the average performance of IA in the network \cite{Stoyan2008,Ganti2009}. Since the underlying point process is stationary, without loss of generality, it can be assumed that the typical transmitter is at the origin\footnote{The whole point process can be translated so the randomly picked node is located at the origin.}. Denote the cluster to which the transmitter at the origin belongs by $\boldsymbol{\Psi}_o$. The received signal at receiver $\hat{x}$, $x\in\boldsymbol{\Psi}_o$, is
\begin{align}
\mathbf{y}_{\hat{x}}&= \sum_{z\in \boldsymbol{\Psi}_o}\sqrt{g_{\hat{x}z}}\mathbf{H}_{\hat{x}z}\mathbf{F}_z\tilde{\mathbf{s}}_z + \boldsymbol{\mathcal{I}}_{\rm{c}} + \mathbf{u}_{\hat{x}},
\label{eq:Clustering_ReceivedSignal}
\end{align}
where $\boldsymbol{\mathcal{I}}_{\rm{c}}= \sum_{z\in \boldsymbol{\Phi}/\boldsymbol{\Psi}_o}\sqrt{g_{\hat{x}z}}\mathbf{H}_{\hat{x}z}\mathbf{F}_z\tilde{\mathbf{s}}_z$ is the inter-cluster interference, $g_{\hat{x}z}$ and $\mathbf{H}_{\hat{x}z}$ represent the pathloss and the matrix of channel coefficients between the transmitter $z$ and the receiver $\hat{x}$, $\mathbf{F}_z$ is the precoder at transmitter $z$ with the transmitted signal $\tilde{\mathbf{s}}_z$ such that $\mathbb{E}\{\tilde{\mathbf{s}}_z^*\tilde{\mathbf{s}}_z\} = P$, and  $\mathbf{u}_{\hat{x}}\sim\mathcal{CN}(\mathbf{0},N_o\mathbf{I})$ is the additive white Gaussian noise. In this paper, it is assumed that $\mathbf{F}_z^*\mathbf{F}_z=\mathbf{I}$, because of tractability and the observation that the gain attained otherwise, such as with the MMSE algorithm in \cite{Peters2011} or the Max-SINR algorithm in \cite{Gomadam2008}, is limited and confined to the low SNR regime where the inter-cluster interference is generally not dominant. In every cluster, channel state information is estimated at the receivers as in \cite{JindalMIMOtraining2010} and conveyed to all other nodes of the cluster using an error-free instantaneous feedback link.
We propose to model the uncertainty in the MIMO channels using a Gauss-Markov model of the form \cite{Wang2007,Musavian2007}
\begin{align}
\mathbf{H}_{\hat{x}z} = \sqrt{1-\beta^2_{\hat{x}z}}\mathbf{H}^w_{\hat{x}z} + \beta_{\hat{x}z}\mathbf{E}_{\hat{x}z}\quad x,z\in\boldsymbol{\Psi}_o, \label{eq:Clustering_ChannelModel}
\end{align}
where $\mathbf{H}^w_{\hat{x}z}$ is the estimated channel, $\mathbf{E}_{\hat{x}z}$ represents the estimation error with i.i.d.\ terms distributed as $\mathcal{CN}(0,1)$, and $\beta_{\hat{x}z}^2$ is the normalized variance of the estimation error. It is assumed that the channel is quasi-static block-fading such that $\mathbf{H}$ is constant for a  block duration of length $T$ and then changes independently. Training, feedback, and data transmission are assumed to be all orthogonal in time, in the same coherence time or frame $T$ \cite{Kobayashi2011}. Hence, $\beta_{\hat{x}z}$ is set to be related to the average received SNR at each link, $\gamma_{\hat{x}z}$, as \cite[Section II.B]{JindalMIMOtraining2010}
\begin{align}
\beta_{\hat{x}z}^2 = \frac{1}{1+\frac{T_{\rm{t}}}{N}\gamma_{\hat{x}z}} = \frac{1}{1+T_{\rm{t}}\frac{\gamma_{\rm{o}}g_{\hat{x}z}}{N}}, \label{eq:Clustering_EstimationVarianceAndSNR}
\end{align}
where $\gamma_{\rm{o}}=\frac{P}{N_{\rm{o}}}$ and $T_{\rm{t}} \geq KN$ is the number of channel instances spent for training $\mathbf{H}_{\hat{x}z}$ \cite{Ayach2012}. For analytical tractability, it is also assumed that $\mathbf{H}^w$ is used to construct the precoders/equalizers and nodes effectively ignore the imperfection in CSI in their design. 

\section{Intra-cluster Interference Alignment} \label{sec:IntraCluster_IA}
\indent At each cluster, a $K$-user system of IA is feasible if there exists a set of matrices \(\mathcal{W}=\{\mathbf{W}_{\hat{z}}| z\in \boldsymbol{\Psi}_0\}\) such that, given the received signal of \eqref{eq:Clustering_ReceivedSignal}, the following constraints are met \cite{Cadambe2008a}:
\begin{align}
\left\{ \begin{array}{l}
{{\rm{rank}}}\left(\mathbf{W}_{\hat{x}}\mathbf{H}_{\hat{x}x}\mathbf{F}_x\right) = N_s \\
\mathbf{W}_{\hat{x}}\mathbf{H}_{\hat{x}z}\mathbf{F}_{z} = \mathbf{0}\ \forall z\neq x
\end{array} \right. \forall x,z\in\boldsymbol{\Psi}_o, \label{eq:IAconds}
\end{align}
where $\mathbf{W}_{\hat{x}}$ is the combining filter used at receiver $\hat{x}$ and $N_s$ is the number of interference-free streams each transmitter can send to its receiver. The linear equalizer presented in \cite{BehrangTSP2011} and the projection matrix presented in \cite[Section III.A]{Peters2011} are examples of a possible receive filter in \eqref{eq:IAconds}. It is assumed that the IA precoders are designed using the alternating minimization algorithm in \cite[Section III.A]{Peters2011} such that $\mathbf{F}_x$ is independent of $\mathbf{H}_{\hat{x}x}$ for all $x\in\boldsymbol{\Psi}$. Also, it is assumed that the set of $\{N, N_s,K\}$ constitutes a feasible IA system, which for the MIMO interference channel requires that  $2N-(K+1)N_s\geq0$.   

\subsection{Characterizing the SINR}
From \eqref{eq:IAconds}, interference at receiver $\hat{x}$ is confined to an $N-N_s$ dimensional subspace. Let $\left[ \{\cdot\}\right]$ represent horizontal concatenation of the elements in $\{\cdot\}$. Then, as IA precoders/equalizers are constructed using $\mathbf{H}^w$ as given by \eqref{eq:Clustering_ChannelModel}, the $N\times(K-1)N_s$ matrix of 
$\mathbf{J}_{\hat{x}}=\left[\{\mathbf{H}^w_{\hat{x}z}\mathbf{F}_z|z\neq x, z\in \boldsymbol{\Psi}_o\}\right]$
spans an $N\!-\!N_s$ dimensional subspace. Let the singular value decomposition of $\mathbf{J}_{\hat{x}}$ be $\mathbf{U}_{\mathbf{J}_{\hat{x}}}\mathbf{\Sigma}_{\mathbf{J}_{\hat{x}}}\mathbf{V}_{\mathbf{J}_{\hat{x}}}^*$ and let the rows of $\mathbf{W}_{\hat{x}}$ be the columns of $\mathbf{U}_{\mathbf{J}_{\hat{x}}}$ corresponding to zero singular values in $\mathbf{\Sigma}_{\mathbf{J}_{\hat{x}}}$. As $\mathbf{W}_{\hat{x}}$ is independent of $\mathbf{H}^w_{\hat{x}x}\mathbf{F}_x$, it satisfies the conditions in \eqref{eq:IAconds} and is a valid zero-forcing (ZF) equalizer for IA. Using this ZF receiver, the post-processing SINR of the $n$th stream at receiver $\hat{x}$ is
\begin{align}
\gamma^{{\rm{IA}}}_{\hat{x},n} \!=\! \frac{g_{\hat{x}x}(1-\beta^2_{\hat{x}x})\tilde{\mathbf{h}}_{\hat{x}x}^*\tilde{\mathbf{h}}_{\hat{x}x}}
{\frac{N_s}{\gamma_{\rm{o}}} + 
\underbrace{g_{\hat{x}x}\beta^2_{\hat{x}x} \tilde{\mathbf{e}}_{\hat{x}x}^*\tilde{\mathbf{e}}_{\hat{x}x}}_{I_s} +  
\underbrace{\sum\limits_{z\in \boldsymbol{\Psi}_o/x}g_{\hat{x}z}\beta^2_{\hat{x}z}\tilde{\mathbf{e}}_{\hat{x}z}^*\tilde{\mathbf{e}}_{\hat{x}z}}_{I_e} +\underbrace{\negthickspace
\sum\limits_{z\in \boldsymbol{\Phi}/\boldsymbol{\Psi}_o}\negthickspace\negthickspace g_{\hat{x}z}\tilde{\mathbf{h}}_{\hat{x}z}^*\tilde{\mathbf{h}}_{\hat{x}z}}_{I_i}},
\label{eq:Clustering_SINR_temp1}
\end{align}
where for all $z\in \boldsymbol{\Phi}$, $\tilde{\mathbf{h}}_{\hat{x}z} \!=\! \left(\mathbf{e}_n^*\mathbf{W}_{\!\hat{x}}\mathbf{H}_{\hat{x}z}^w\mathbf{F}_{\!z}\right)^*$, 
$\tilde{\mathbf{e}}_{\hat{x}z} \!=\! \left(\mathbf{e}_n^*\mathbf{W}_{\!\hat{x}}\mathbf{E}_{\hat{x}z}\mathbf{F}_{\!z}\right)^*$, and $\mathbf{e}_n$ is the $n$th column of an $N_s\times N_s$ identity matrix. Note that $I_s$ represents the residual error from the direct link and, as the distance between the transmitter and receiver is constant, its pathloss (and the error variance) are not random variables and so it is separated from $I_e$ to emphasize this point.
Let the entries of $\mathbf{H}_{\hat{x}z}$ and $\mathbf{E}_{\hat{x}z}$ be i.i.d.\ Gaussian terms. As $\mathbf{W}_{ \hat{x}}$ and $\mathbf{F}_{ z}$ are unitary matrices independent of $\mathbf{H}_{\hat{x}z}$ and $\mathbf{E}_{\hat{x}z}$, due to the doubly unitarily invariance of the Gaussian distribution, $\tilde{\mathbf{h}}_{\hat{x}z}$ and $\tilde{\mathbf{e}}_{\hat{x}z}$ will be column vectors of length $N_s$ with i.i.d.\ Gaussian terms and independent of each other (as $\mathbf{H}_{\hat{x}z}$ and $\mathbf{E}_{\hat{x}z}$ are independent of each other). Note that \eqref{eq:Clustering_SINR_temp1} is in fact independent of the stream index $n$.

\subsection{Probability of Outage}
In \eqref{eq:Clustering_SINR_temp1}, since  $\tilde{\mathbf{h}}_{\hat{x}z}^*\tilde{\mathbf{h}}_{\hat{x}z}$ and $\tilde{\mathbf{e}}_{\hat{x}z}^*\tilde{\mathbf{e}}_{\hat{x}z}$ are i.i.d.\ $\Gamma(N_s,1)$ random variables, we denote them both by $h_{\hat{x}z}$ for notational simplicity. Denote the typical transmitter at the origin by $o$. Considering a transmitter at the origin is equivalent to conditioning on the existence of a point at the origin. Since every point belongs to some cluster, conditioning on the existence of a point at the origin equals the presence of a cluster with a daughter point at the origin. Since the parent point process is a PPP, an additional cluster with a daughter point at the origin can be added to it without changing the statistics of the other points of the process. Succinctly, the Palm probability of a Neyman-Scott cluster process is $\P^{o}=\P*\boldsymbol{\Psi}_o$ where $*$ denotes  superposition \cite{Stoyan2008}.  This implies that assuming a point of the cluster process at the origin equals the original point process $\boldsymbol{\Phi}$ plus an additional cluster which has a point at the origin. Also this additional cluster at the origin $\boldsymbol{\Psi}_o$ is independent of the original process $\boldsymbol{\Phi}$.  Since the tagged transmitter at origin  does not contribute to the interference at the receiver, it is convenient to use the reduced Palm probability denoted by $\PP$ instead of Palm probability. Reduced Palm probability is similar to Palm probability, except that the point at the origin is not considered in the computation of the probability and hence $\PP=\P*\{\boldsymbol{\Psi}_o\setminus \{o\}\}$. From \eqref{eq:Clustering_SINR_temp1}, the probability of success is therefore
$
{\rm{P_s^{IA}}}(\theta)=\PP(\gamma^{{\rm{IA}}}_{\hat{o},n}>\T)
$, where $\theta$ is the SINR threshold. 

\begin{theorem}
\label{thm:one}
For the system model described in Section \ref{sec:system_model}, the success probability when each cluster uses IA is
\begin{align}
{\rm{P_s^{IA}}}(\theta) =  \sum_{k=0}^{N_s-1}\! \frac{(-\eta)^k}{k!} \frac{\d^k}{\d s^k}e^{- s\frac{N_s}{\gamma_{\rm{o}}}}
\left(\frac{\frac{T_{\rm{t}}\gamma_{\rm{o}}}{N}+D_{\rm{r}}^\alpha}{s+\frac{T_{\rm{t}}\gamma_{\rm{o}}}{N}+D_{\rm{r}}^\alpha}\right)^{N_s}
\L^{!o}_{\I_e}\!(s)\L_{\I_i}\!(s)\Big{|}_{s=\eta}, \label{eq:ProbSuccessIA_TheoEq}
\end{align}
where $\L^{!o}_{\I_e}(s)$ and $\L_{\I_i}(s)$ are Laplace transforms of intra and inter-cluster interference given by
\begin{align}
\L^{!o}_{\I_e}(s)&= 
\frac{1}{\pi R^2}\int\limits_{\!\!B(o,R)}
\negmedspace
\Biggg[\!\frac{1}{\pi\! R^2}
\negthickspace\!
\int\limits_{\!\!B(o,R)} \negthickspace\negthickspace\negmedspace\left(\!\frac{\frac{T_{\rm{t}}\gamma_{\rm{o}}}{N}\!+\!\|x\!-\!y\!-\!\hat{o}\|^\alpha}{s\!+\!\frac{T_{\rm{t}}\gamma_{\rm{o}}}{N}\!+\!\|x\!-\!y\!-\!\hat{o}\|^\alpha}\!\right)^{\!\!\!N_s}
\negmedspace\negmedspace
\d x\Biggg]^{\!K\!-\!1}
\negthickspace\negthickspace\negthickspace\!
\d y,
\label{eq:lap1}
\\
\L_{\I_i}(s)&=\!\exp\!\Bigg(
\negthickspace\negthickspace
-\lambda_p \! \int\limits_{\R^2}
\negthickspace\!
1\negmedspace-\!\negmedspace\Bigg[\!\frac{1}{\pi\! R^2}
\negthickspace\!\!
\int\limits_{B(o,R)}
\negthickspace\negthickspace\negmedspace\!
\left(\!\frac{\|x\!-\!y\|^{\alpha}}{s\!+\!\|x\!-\!y\|^{\alpha}}\!\!\right)^{\negthickspace\! N_s}
\negthickspace\negthickspace\negmedspace
\d x\Bigg]^{\!K}
\negthickspace
\d y
\Bigg), \label{eq:lap2}
\end{align}
where 
$\eta= \frac{\T}{g_{\hat{o}o} (1-\beta^2_{\hat{o}o})}$.
\end{theorem}
\begin{IEEEproof}
See Appendix \ref{AppendixProof:PoutTheorem}.
\end{IEEEproof}
Note that the integrals in \eqref{eq:lap1} and \eqref{eq:lap2} can be computed by switching to polar coordinates and exchanging the order of differentiation and integration\footnote{The integral in \eqref{eq:lap1} is always finite since the domain of integration is a bounded set.  The integral is \eqref{eq:lap2} can be shown to be finite using the fact that $\|x-y\| \geq \|y\| - R$ for the inner integral and $1-x^m\leq \exp(-x^m),\  m\geq 1$ for the outer integral.}. Although the probability of successful transmission given by \eqref{eq:ProbSuccessIA_TheoEq} is in closed-form, numerically computing it, especially for $N_s>1$ where differentiation is required, is not trivial. Lemma \ref{lem:UpperBoundForThePoutTerms} provides bounds on the Laplace transforms in \eqref{eq:lap1} and \eqref{eq:lap2} that are easily computable. Also, Lemma \ref{lem:UpperBoundOnPs_noDerivitive} can be used to avoid the differentiation in \eqref{eq:ProbSuccessIA_TheoEq} when $N_s>1$.

\begin{lem} \label{lem:UpperBoundForThePoutTerms}
For $K>2$
\begin{align}
\L_{\I_i}(s) &\leq \exp\left(-\lambda_p s^{2/\alpha}\frac{\Gamma(KN_s+2/\alpha)\Gamma(1-2/\alpha)}{\Gamma(KN_s)} \right)
\label{eq:L_i_UpperBound}
\\
\L^{!o}_{\I_e}(s) &\leq \frac{1}{2\pi} \int_{\theta=0}^{2\pi}\int_{x=0}^{2R} \frac{h_r(x)}{(1+sf(\sqrt{x^2+D_{\rm{r}}^2-2xD_r\cos(\theta)}))^{N_s(K-1)}}\d x \d \theta,
\label{eq:L_Ie_UpperBound}
\end{align}
where
\begin{align}
h_r(x)=\left\{\begin{array}{ll} \frac{x}{R^2}I_{1-\frac{x^2}{4R^2}}(3/2,1/2)& x\leq 2R
\\0 & \text{\rm{Otherwise}}\end{array}\right.  \label{eq:PDFofR_Uniform}
\end{align}
and $I_y(a,b)$ is the regularized incomplete beta function.
\end{lem}
\begin{IEEEproof}
See Appendix \ref{AppendixProof:UpperBoundForThePoutTerms}.
\end{IEEEproof}

\begin{lem} \label{lem:UpperBoundOnPs_noDerivitive}
If $\eta>(N_s-1)/e$,  the success probability is bounded by
\begin{align}
{\rm{P_s^{IA}}} \leq \sum_{k=0}^{N_s-1} \frac{\eta^k}{k!}e^{-(\eta-k/e)\frac{N_s}{\gamma_{\rm{o}}}}
\E e^{-(\eta - k/e) g_{\hat{o}o}\beta^2_{\hat{o}o} h_{\hat{o}o} }
\L^{!o}_{\I_e}(\eta-k/e)\L_{\I_i}(\eta-k/e). \label{eq:Lemma2_firstUpperBoundOnPs}
\end{align}
\end{lem}
\begin{IEEEproof}
See Appendix \ref{AppendixProof:UpperBoundOnPs_noDerivitive}.
\end{IEEEproof}
For $N_s>1$, unlike \eqref{eq:ProbSuccessIA_TheoEq}, \eqref{eq:Lemma2_firstUpperBoundOnPs} does not have a differentiation operator. Note that the condition on $N_s$ in Lemma \ref{lem:UpperBoundOnPs_noDerivitive} is equivalent to $N_s<1+\frac{e\theta D_{\rm{r}}^\alpha}{1-\beta_{\hat{o}o}^2}$ which is true for practical $N_s$ and typical operating regime where $\theta\gg 1$, $D_r>1$, and $\beta_{\hat{o}o}^2\ll 1$. Also, results from Lemma \ref{lem:UpperBoundForThePoutTerms} can be used to further simplify the expression in Lemma \ref{lem:UpperBoundOnPs_noDerivitive}.

\subsection{Optimizing the Training Period}
For a given block fading of length $T$, the transmitters spend $T_{\rm{t}}\geq KN$ channel instances for training the links. We also assume a prefect analog feedback link where the receivers send the trained channels over a period of $T_{{\rm{f}}} = K^2N$ channel instances to the transmitters. In practice, the transmitters select $T_{\rm{t}}$ to optimize some performance criteria. In this paper, we assume transmitters use the goodput at each cluster defined as 
\begin{align}
\begin{split}
\hat{T}_{{\rm{t}}} = \operatorname*{arg\,max}_{T_{\rm{t}}} & \quad \frac{T-K^2N - T_{\rm{t}}}{T} KN_s {\rm{P_s^{IA}}}(\theta)\log_2(1+\theta) 
\\
{\rm{s.t.}} \quad & KN\leq T_{\rm{t}}\leq T-K^2N, 
\end{split}
\label{eq:trainingOptimization}
\end{align}
where $\frac{T-K^2N - T_{\rm{t}}}{T}$ accounts for the transmission opportunities lost due to overhead, $KN_s$ is the total number of streams in each cluster, and ${\rm{P_s^{IA}}}(\theta)\log_2(1+\theta)$ is the rate multiplied by the times the \textit{connection exists}, i.e. SINR passes the threshold $\theta$. Note that in \eqref{eq:trainingOptimization}, ${\rm{P_s^{IA}}}(\theta)$ is implicitly a function of the training period $T_{\rm{t}}$.

\indent For a given node mobility and hence a Doppler frequency $f_d\approx \frac{1}{T}$, the training period can be written as a fraction of the total block length $T$, i.e. $T_{\rm{t}} = \delta T = \delta \frac{1}{f_d}$. Therefore, the optimization problem of \eqref{eq:trainingOptimization} can be rewritten as
\begin{align}
\begin{split}
\hat{\delta} = \operatorname*{arg\,max}_{\delta} &\quad (1-\delta-f_dK^2N) K N_s {\rm{P_s^{IA}}}(\theta)\log_2(1+\theta) 
\\
{\rm{s.t.}} \quad & f_d K N \leq\delta\leq (1-f_d K^2 N) ,
\end{split}
\label{eq:TrainingOpt_Ns1_fdFormulation_1}
\end{align}
where 
\begin{align}
{\rm{P_s^{IA}}}=\sum_{k=0}^{N_s-1} \frac{(-\eta)^k}{k!} \frac{\d^k}{\d s^k}e^{- s\frac{N_s}{\gamma_{\rm{o}}}}\! 
\left(\frac{\frac{\delta \gamma_{\rm{o}}}{N}+f_dD_r^\alpha}{\frac{\delta \gamma_{\rm{o}}}{N}+f_d(s+D_{\rm{r}}^\alpha)}\right)^{N_s}
\L^{!o}_{\I_e}\!(s,f_d)
\L_{\I_{i}}\!(s)\Big{|}_{\!s=\eta}, \nonumber
\end{align}
$\eta= \frac{\theta {D_{r}}^\alpha \left(\gamma_{\rm{o}}\delta + N {D_{r}}^\alpha f_d\right)}{\gamma_{\rm{o}} \delta}$,
$\L^{!o}_{\I_e}\!(s,f_d) = \frac{1}{\pi R^2}\negthickspace\int\limits_{\!B(o,R)}
\negthickspace
\left[\!\frac{1}{\pi\! R^2}
\negthickspace\negthickspace\!
\int\limits_{\!B(o,R)} \negthickspace\frac{\frac{\delta \gamma_{\rm{o}}}{N}+ f_d\|x-y-\hat{o}\|^\alpha}{\frac{\delta \gamma_{\rm{o}}}{N}+f_d(s+\|x-y-\hat{o}\|^\alpha)}
\d x\right]^{\!K-1}
\negthickspace\negthickspace\negthickspace
\d y$
, and $\L_{\I_{i}}\!(s)$ is given by \eqref{eq:lap2}.

\indent With a convex relaxation on $T_{\rm{t}}$ (and therefore $\delta$) to change its domain to the real numbers, the optimization problem of \eqref{eq:TrainingOpt_Ns1_fdFormulation_1} is convex and solvable with any of the numerical optimization algorithms \cite{boyd2004convex}. Note that, although complicated, the derivative of the objective function in \eqref{eq:TrainingOpt_Ns1_fdFormulation_1} w.r.t $\delta$ is computable and evaluating the objective function or its derivatives for any set of values is possible. Next we provide approximate closed form solutions for some common cases.

\subsubsection{Single stream from each transmitter} \label{subsec:singleStreamTrainingPeriorOptimization}
When $N_s=1$, ${\rm{P_s^{IA}}}$ in \eqref{eq:TrainingOpt_Ns1_fdFormulation_1} simplifies to
\begin{align}
{\rm{P_s^{IA}}}(N_s=1)= e^{- \eta\frac{N_s}{\gamma_{\rm{o}}}}\! 
\frac{\frac{\delta \gamma_{\rm{o}}}{N}+f_dD_r^\alpha}{\frac{\delta \gamma_{\rm{o}}}{N}+f_d(\eta+D_{\rm{r}}^\alpha)}
\L^{!o}_{\I_e}\!(\eta,f_d)
\L_{\I_{i}}\!(\eta). \label{eq:Ns_1_FullOptimization}
\end{align}
The highly non-linear dependency of \eqref{eq:Ns_1_FullOptimization} on $\delta$ can be converted into a polynomial one following the Taylor expansion approximation method proposed in \cite{Jindal2010_pilot_taylor}. Let $g(\delta,f_d)$ be the objective function in \eqref{eq:TrainingOpt_Ns1_fdFormulation_1}. Rewrite $g(\delta,f_d)$ as its Taylor series around $f_d=0$ (infinite block length and hence perfect training) keeping all the other variables constant 
\begin{align}
\begin{split}
\hat{\delta} \approx \operatorname*{arg\,max}_{\delta} &\quad g(\delta,f_d=0) + \frac{\partial g}{\partial f_d}|_{f_d=0}f_d + \frac{1}{2}\frac{\partial^2 g}{\partial f_d^2}|_{f_d=0}f_d^2 
\\
{\rm{s.t.}} \quad & f_d KN\leq\delta\leq (1-f_dK^2N). \end{split}
\label{eq:TrainingOpt_Ns1_fdFormulation_2}
\end{align}
Simplifying the terms in \eqref{eq:TrainingOpt_Ns1_fdFormulation_2} and removing the constant scaling coefficients from the optimization problem yields
\begin{align}
\begin{split}
\hat{\delta}_1 \approx \operatorname*{argmax}_{\delta} &\quad 
- \delta + C_1 \frac{1}{\delta} + C_2 \frac{1}{\delta^2}
\\
{\rm{s.t.}} \quad & f_d KN\leq\delta\leq (1-f_dK^2N), \end{split}
\label{eq:TrainingOpt_Ns1_fdFormulation_3}
\end{align}
where  $C_1$ and $C_2$ are given in Appendix \ref{Appendix:C1_and_C2}. Let $\delta_1$ be the relevant root of the first derivative of \eqref{eq:TrainingOpt_Ns1_fdFormulation_3}. The optimum training period will be given by
\begin{align}
\hat{T}_{t,1} = \min\left(\max\left(KN,[\delta_1 T]\right), T - K^2N\right). \label{eq:delta_hat_1}
\end{align}

\indent When $\frac{T_{\rm{t}}}{N}g_{\hat{o}o}\gamma_{\rm{o}} = \frac{T_{\rm{t}}}{N}\frac{\gamma_{\rm{o}}}{D_r^{\alpha}}\gg 1$, which is true for high SNR, $\beta_{\hat{o}o}\approx 0$ which implies that $\eta\approx \theta D_{\rm{r}}^{\alpha}$. In this case, $\L_{\I_i}\!(\eta)$ is a strictly positive function independent of $T_{\rm{t}}$ and \eqref{eq:TrainingOpt_Ns1_fdFormulation_3} simplifies to 
\begin{align}
\begin{split}
\hat{\delta}_2 \approx \operatorname*{argmax}_{\delta} &\quad 
- \delta + D_1 \frac{1}{\delta} + D_2 \frac{1}{\delta^2}
\\
{\rm{s.t.}} \quad & f_d KN\leq\delta\leq (1-f_dK^2N), \end{split}
\label{eq:TrainingOpt_Ns1_fdFormulation_simple_1}
\end{align}
where $D_1$ and $D_2$ are given in Appendix \ref{Appendix:D1_and_D2}. Similar to $\hat{T}_{t,1}$, let $\delta_2$ be the relevant root of the first derivative of \eqref{eq:TrainingOpt_Ns1_fdFormulation_simple_1}. The optimum training period will be given by
\begin{align}
\hat{T}_{t,2} = \min\left(\max\left(KN,[\delta_2 T]\right), T - K^2N\right). \label{eq:delta_hat_2}
\end{align}

\subsubsection{Greater than one stream from each transmitter} \label{subsubsec:GreaterThanOneStreamOptimumTraining}
\indent For $N_s>1$, the same approach taken to derive \eqref{eq:delta_hat_1} and \eqref{eq:delta_hat_2} can be used to approximately solve for the optimum training period. Alternatively, bounds given in \eqref{eq:L_i_UpperBound}, \eqref{eq:L_Ie_UpperBound},  and Lemma \ref{lem:UpperBoundOnPs_noDerivitive} can be used to derive simpler (and less accurate) results. We believe the involved expressions in this case do not provide additional insight into the problem at this point and are not presented here.

\subsection{Transmission Capacity}
Let $q(\lambda_p) = 1-{\rm{P_s^{IA}}}(\theta)=\epsilon$. Accounting for overhead, the normalized transmission capacity is 
\begin{align}
C(\epsilon) = \frac{T - K^2N - \hat{T}_{{\rm{t}}}}{T}q^{-1}(\epsilon)K(1-\epsilon). \label{eq:TransmissionCapacityIA}
\end{align}
With the probability of successful transmission as given by \eqref{eq:ProbSuccessIA_TheoEq}, the exact expression of $q^{-1}(\epsilon)$ for general $N_s$ is not analytically tractable. Next we give its exact expression for the case of $N_s=1$ and provide a bound for general $N_s>1$.
\subsubsection{Single stream from each transmitter}
Using \eqref{eq:ProbSuccessIA_TheoEq}
\begin{align}
q(\lambda_p) = 1 - e^{- \hat{\eta}\frac{N_s}{\gamma_{\rm{o}}}}\! \left(\frac{\frac{\hat{T}_{\rm{t}}\gamma_{\rm{o}}}{N}+D_{\rm{r}}^\alpha}{s+\frac{\hat{T}_{\rm{t}}\gamma_{\rm{o}}}{N}+D_{\rm{r}}^\alpha}\right)\L^{!o}_{\I_{e}}\!(\hat{\eta})\L_{\I_{i}}\!(\hat{\eta}) = \epsilon, \label{eq:TransCap_1}
\end{align}
where $\hat{\eta} = \eta|_{T_{{\rm{t}}} = \hat{T}_{{\rm{t}}}}$. Substituting \eqref{eq:lap1} and \eqref{eq:lap2} into \eqref{eq:TransCap_1} gives
\begin{align}
q^{-1}(\epsilon) &= \lambda_p^{\epsilon} 
= \frac{\log_e\left(
\frac{e^{- \hat{\eta}\frac{N_s}{\gamma_{\rm{o}}}}}{1-\epsilon} 
\frac{\frac{\hat{T}_{{\rm{t}}} \gamma_{\rm{o}}}{N}+D_{\rm{r}}^\alpha}{\frac{\hat{T}_{{\rm{t}}} \gamma_{\rm{o}}}{N}+\hat{\eta}+D_{\rm{r}}^\alpha}
\frac{1}{\pi R^2}\int\limits_{\!\!B(o,R)}
\negthickspace\negmedspace
\Biggg[\!\frac{1}{\pi\! R^2}
\negthickspace\negthickspace\!
\int\limits_{\!\!B(o,R)} \negthickspace\negthickspace\negmedspace\left(\!\frac{\frac{\hat{T}_t \gamma_{\rm{o}}}{N }+\|x-y-\hat{o}\|^\alpha}{\hat{\eta}+ \frac{\hat{T}_{{\rm{t}}} \gamma_{\rm{o}}}{N}+\|x-y-\hat{o}\|^\alpha}\!\right)^{\!\!\!N_s}
\negthickspace\negmedspace\negmedspace
\d x\Biggg]^{\!K\!-\!1}
\negthickspace\negthickspace\negthickspace\negthickspace\!
\d y.
\right)}
{\negmedspace\! \int\limits_{\R^2}
\negthickspace
1\negmedspace-\!\negmedspace\Bigg[\!\frac{1}{\pi\! R^2}
\negthickspace\negthickspace
\int\limits_{B(o,R)}
\negthickspace\negmedspace\!
\left(\!\frac{\|x-y\|^{\alpha}}{\hat{\eta}+\|x-y\|^{\alpha}}\!\!\right)^{\negthickspace N_s}
\negthickspace
\d x\!\Bigg]^{\!K}
\negthickspace\negthickspace
\d y}, \label{eq:Q_inverse_epsilon_Ns_1_IA}
\end{align}
where $\hat{T}_{{\rm{t}}}$ is the optimum training period obtained earlier. Using \eqref{eq:Q_inverse_epsilon_Ns_1_IA}, \eqref{eq:TransmissionCapacityIA} can be computed.

\subsubsection{Greater than one stream from each transmitter} 
When $\hat{\eta} > (N_s-1)/e$, using Lemma \ref{lem:UpperBoundOnPs_noDerivitive}
\begin{align}
q(\lambda_p) \geq 1- \sum_{k=0}^{N_s-1} \frac{\hat{\eta}^k}{k!}e^{-(\hat{\eta}-k/e)\frac{N_s}{\gamma_{\rm{o}}}}
\E e^{-(\hat{\eta}-k/e) g_{\hat{o}o}\beta^2_{\hat{o}o} h_{\hat{o}o} }
\L^{!o}_{\I_e}(\hat{\eta}-k/e)\L_{\I_i}(\hat{\eta}-k/e). 
\label{eq:temp6}
\end{align}
Note that the expression for the optimum training when $N_s>1$ was not presented in Section \ref{subsubsec:GreaterThanOneStreamOptimumTraining}.
Now, if $\hat{\eta} \gg k/e$ for all $k\in\{0,\ldots,N_s-1\}$, $\hat{\eta}-k/e \approx \hat{\eta}$ and hence \eqref{eq:temp6} equals
\begin{align}
q(\lambda_p) &\geq 1- \left(\sum_{k=0}^{N_s-1} \frac{\hat{\eta}^k}{k!}\right)e^{-\hat{\eta}\frac{N_s}{\gamma_{\rm{o}}}}
\E e^{-\hat{\eta} g_{\hat{o}o}\beta^2_{\hat{o}o} h_{\hat{o}o} }
\L^{!o}_{\I_e}(\hat{\eta})\L_{\I_i}(\hat{\eta})
\nonumber \\
\Rightarrow  \L_{\I_i}(\hat{\eta}) &\geq \frac{1-\epsilon}
{\left(\sum_{k=0}^{N_s-1} \frac{\hat{\eta}^k}{k!}\right)e^{-\hat{\eta}\frac{N_s}{\gamma_{\rm{o}}}}
\E e^{-\hat{\eta} g_{\hat{o}o}\beta^2_{\hat{o}o} h_{\hat{o}o} }
\L^{!o}_{\I_e}(\hat{\eta}) }
\nonumber \\
\Rightarrow  \lambda_p^\epsilon &\leq \frac{\log_e\left(
\frac{1}{1-\epsilon}
\left(\sum_{k=0}^{N_s-1} \frac{\hat{\eta}^k}{k!}\right)e^{-\hat{\eta}\frac{N_s}{\gamma_{\rm{o}}}}
\E e^{-\hat{\eta} g_{\hat{o}o}\beta^2_{\hat{o}o} h_{\hat{o}o} }
\L^{!o}_{\I_e}(\hat{\eta}) 
\right)}
{\negmedspace\! \int\limits_{\R^2}
\negthickspace
1\negmedspace-\!\negmedspace\Bigg[\!\frac{1}{\pi\! R^2}
\negthickspace\negthickspace
\int\limits_{B(o,R)}
\negthickspace\negmedspace\!
\left(\!\frac{\|x-y\|^{\alpha}}{\hat{\eta}+\|x-y\|^{\alpha}}\!\!\right)^{\negthickspace N_s}
\negthickspace
\d x\!\Bigg]^{\!K}
\negthickspace\negthickspace
\d y }.
 \label{eq:BoundOnTransCap_2}
\end{align}
To simplify the integrations in \eqref{eq:BoundOnTransCap_2}, one could also use the bound given by \eqref{eq:L_Ie_UpperBound} for $\L^{!o}_{\I_e}(\hat{\eta})$ or use a similar approach taken to obtain \eqref{eq:L_i_UpperBound} to find an appropriate bound for the denominator of \eqref{eq:BoundOnTransCap_2}.

\section{Spatial Multiplexing} \label{sec:SM}
An alternative strategy to IA is spatial multiplexing with an orthogonal multiple access technique, e.g. TDMA. We analyze open-loop spatial multiplexing where at each cluster only a single transmit/receiver pair communicate $N$ streams without precoding. In short, we refer to TDMA+SM as SM. For a fair comparison with IA, we assume a similar CSI imperfection as in \eqref{eq:Clustering_ChannelModel}. Therefore, the signal at a typical receiver can be written as
\begin{align}
\mathbf{y}_{\hat{x}}\! =\!g_{\hat{x}x}\left(\sqrt{(1\!-\!\beta_{\hat{x}x}^2)g_{\hat{x}x}}\mathbf{H}^w_{\hat{x}x}\!+\!\beta_{\hat{x}x}\mathbf{E}_{\hat{x}x}\right)\tilde{\mathbf{s}}_{x} \!+\! \boldsymbol{\mathcal{I}}_c \!+\! \mathbf{u}_{\hat{x}},
\nonumber 
\end{align}
where $\boldsymbol{\mathcal{I}}_c$ is defined in \eqref{eq:Clustering_ReceivedSignal} with the difference that each cluster only has a single transmitter. In this case, the point process of the transmitters simplifies to the Poisson point process of the parent points. When $\beta_{\hat{x}x}=0$, after a zero-forcing receiver based on $\mathbf{H}_{\hat{x}x}$, the SINR of the $n$th stream at a typical receiver can be written as
\begin{align}
\gamma^{\rm{SM}}_{\hat{x},n} = \frac{g_{\hat{x}x}}
{\mathbf{e}_n^*
\left(\mathbf{H}_{\hat{x}x}^w\right)^{-1}
\left(
\frac{N}{\gamma_{\rm{o}}}\mathbf{I}_N +
\sum\limits_{z\in \boldsymbol{\Phi}/\boldsymbol{\Psi}_o}
g_{\hat{x}z}\mathbf{H}_{\hat{x}z}\mathbf{H}_{\hat{x}z}^*\right)
\left(\left(\mathbf{H}_{\hat{x}x}^w\right)^{-1}\right)^*
\mathbf{e}_n},	
\label{eq:SM_ZF_SINR}
\end{align}
where we have assumed the interference values from different transmitters are independent. The probability of success is given by \cite{Louie2011}
\begin{align}
\rm{P_s^{SM}}(\theta)&=\mathbb{P}^{!o}(\gamma_{\hat{x},n}^{\rm{SM}}>\theta) 
=\exp\left(-\lambda_p
\theta^{\frac{2}{\alpha}} D_{\rm{r}}^{2}\mathcal{J}-\theta D_{\rm{r}}^{\alpha}\frac{N}{\gamma_{\rm{o}}}
\right),
\label{eq:Psuccess_SM_ZF}
\end{align}
where $\mathcal{J}=\frac{\pi \Gamma\left(N+\frac{2}{\alpha}\right)\Gamma\left(1-\frac{2}{\alpha}\right)}{\Gamma\left(N\right)}$. With imperfect CSI, we assume the receivers compute their zero-forcing receivers based on the estimated channel values and ignore the estimation error. Then, the denominator of \eqref{eq:SM_ZF_SINR} will have an additional term of $g_{\hat{x}x}\beta^2_{\hat{x}x}\mathbf{E}_{\hat{x}x}\mathbf{E}_{\hat{x}x}^*$ inside the parenthesis. As the channel estimation error is independent of the estimated channel and the transmitter/receiver distance is fixed to $D_{\rm{r}}$, this additional term is effectively increasing the noise spectral density from $N_{\rm{o}}$ to $N_{\rm{o}} + Pg_{\hat{x}x}\beta^2_{\hat{x}x}$. As the numerator of \eqref{eq:SM_ZF_SINR} also changes to $g_{\hat{x}x}\left(1-\beta^2_{\hat{x}x}\right)$, in case of imperfect CSI as in \eqref{eq:Clustering_ChannelModel}, the probability of success at a typical receiver for any of the streams changes from \eqref{eq:Psuccess_SM_ZF} to
\begin{align}
\rm{P_s^{SM}}(\theta)&=\mathbb{P}^{!o}(\gamma_{\hat{x},n}^{\rm{SM}}>\theta) 
=\exp\left(-\lambda_p
\tilde{\theta}^{\frac{2}{\alpha}} D_{\rm{r}}^{2}\mathcal{J}-\tilde{\theta} D_{\rm{r}}^{\alpha}\frac{N\tilde{N}_o}{P}
\right),
\label{eq:Psuccess_SM_ZF_imperfectCSI}
\end{align} 
where $\tilde{\theta} = \frac{\theta}{1-\beta^2_{\hat{x}x}}$ and $\tilde{N}_o=N_{\rm{o}}+Pg_{\hat{x}x}\beta^2_{\hat{x}x}$.

\subsection{Optimum Training Period}
Similar to the IA case, each node selects a training period that optimizes its own (its own cluster's) goodput. Unlike IA, however, here the minimum training period for each transmit/receive pair is $N$ and no resources are spent for feedback. Hence
\begin{align}
\begin{split}
\hat{T}_{{\rm{t}}}^{{\rm{SM}}} = \operatorname*{argmax}_{T_{\rm{t}}} & \quad \frac{T - T_{\rm{t}}}{T} N {\rm{P_s^{SM}}}(\theta)\log_2(1+\theta) 
\\
{\rm{s.t.}} \quad & N\leq T_{\rm{t}}\leq T .
\end{split}
\label{eq:trainingOptimization_SM}
\end{align}
It is possible to show that the optimization problem of \eqref{eq:trainingOptimization_SM}, after a convex relaxation on $T_{\rm{t}}$, is also a convex problem and any of the numerical optimization algorithms \cite{boyd2004convex} can be used to solve for the optimum $\hat{T}_{{\rm{t}}}^{{\rm{SM}}}$. Nevertheless, similar to the optimization problem of \eqref{eq:TrainingOpt_Ns1_fdFormulation_1}, we first replace the objective function of \eqref{eq:trainingOptimization_SM} with its second order Taylor expansion around $f_d=0$ (infinite block length and hence perfect training) holding all the other variables constant, and then obtain an approximate closed-form solution for $\hat{T}_{{\rm{t}}}^{{\rm{SM}}}$
\begin{align}
\begin{split}
\hat{\delta}^{{\rm{SM}}} = \operatorname*{argmax}_{\delta^{{\rm{SM}}}} & \quad (1-\delta) N {\rm{P_s^{SM}}}(\theta)\log_2(1+\theta) 
\\
{\rm{s.t.}} \quad & N\leq T_{\rm{t}}\leq T .
\end{split}
\label{eq:trainingOptimization_SM_fdBased_1}
\end{align}
Removing the constant coefficients and simplifying the terms, the new optimization problem derived from the 2nd order Taylor series expansion of the objective function in \eqref{eq:trainingOptimization_SM_fdBased_1} is
\begin{align}
\begin{split}
\hat{\delta}^{{\rm{SM}}} \approx \operatorname*{argmax}_{\delta} &\quad 
- \delta + B_1 \frac{1}{\delta} + B_2 \frac{1}{\delta^2}
\\
{\rm{s.t.}} \quad & f_d N\leq\delta\leq 1, 
\end{split}
\label{eq:TrainingOpt_SM_fdFormulation_1}
\end{align}
where $B_1$ and $B_2$ are given in Appendix \ref{Appendix:B1_and_B2}. Similar to the IA case, let $\delta_3$ be the relevant root of the first derivative of \eqref{eq:TrainingOpt_SM_fdFormulation_1}. The optimum training period will be given by
\begin{align}
\hat{T}_{t}^{{\rm{SM}}} = \min\left(\max\left(N,[\delta_3 T]\right), T\right). \label{eq:delta_hat_SM}
\end{align}

\subsection{Transmission Capacity}
For a given block length $T$, accounting for overhead, the normalized transmission capacity is
\begin{align}
c(\epsilon) = \frac{T-\hat{T}_{t}^{{\rm{SM}}}}{T}N\lambda_p^{\epsilon}(1-\epsilon).  \label{eq:TC_SM}
\end{align}
Using \eqref{eq:Psuccess_SM_ZF_imperfectCSI}, $\lambda_p^{\epsilon}$ is found as
\begin{align}
\lambda_p^{\epsilon} = \frac{N}{\tilde{\theta}^{2/\alpha}D_{\rm{r}}^2\mathcal{J}}
 \left(\log\left(\frac{1}{1-\epsilon}\right) - \frac{\tilde{\theta} D_{\rm{r}}^{\alpha}N\tilde{N}_o}{P} \right), \label{eq:Q_inverse_SM}
\end{align}
where in computing $\tilde{\theta}$ and $\tilde{N}_o$, the optimum training period, $\hat{T}_{{\rm{t}}}^{{\rm{SM}}}$, is used.

\section{Numerical Examples and Discussion} \label{sec:NumericalResults}
In this section, we present numerical examples demonstrating both the accuracy of the developed theory for analyzing large IA networks and the effectiveness of such analysis in discovering operating regimes where IA outperforms other transmission schemes. For simplicity, we focus on the case of $K\!=\!3$, $N\!=\!2$, and $N_s\!=\!1$. To further reduce the number of involved variables, we set $R$ to $D_c\! =\! 0.5\lambda_p^{0.5}$ (average distance between the cluster centers) and set $D_{\rm{r}}$ to $\frac{R}{5}$. In this way, by increasing $D_c$, all the nodes move away from each other. We set $N_{\rm{o}}$ to $1$ and unless otherwise stated, it is also assumed that all the clusters are transmitting simultaneously.
In the forthcoming discussions, perfect channel estimation corresponds to very large values of $T_{\rm{t}}$ where $\beta$ is assumed to be $0$ and the worst channel estimation corresponds to assigning the least number of channel instances required for training the cross links in an IA cluster, i.e. $T_{\rm{t}}=KN$. Note that for perfect CSI, \eqref{eq:coeff_for_imperfectCSI} and \eqref{eq:lap1} are both equal to $1$.

\begin{figure}
\centering
\includegraphics[width=4.3in]{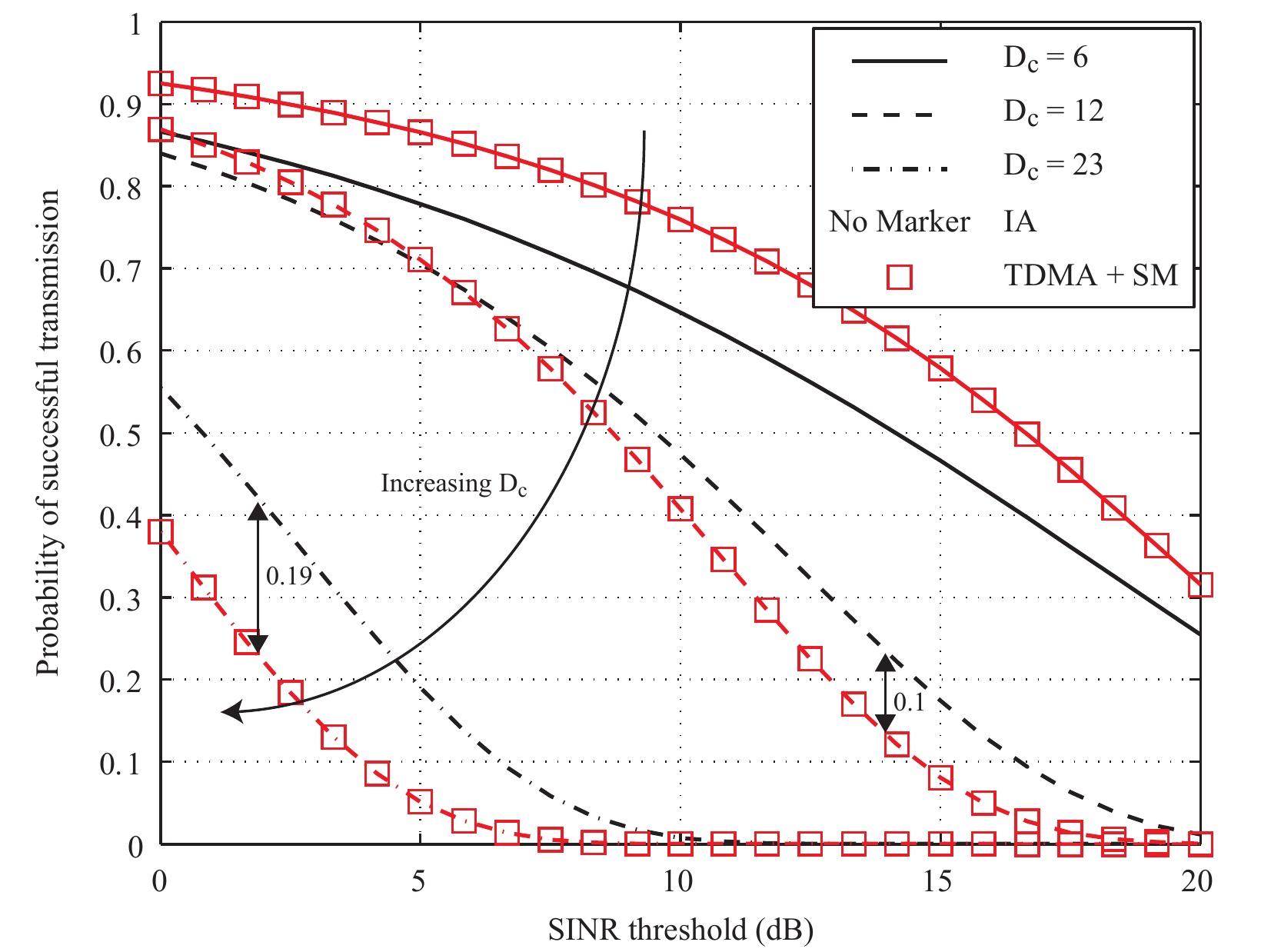}
\caption{The probability of successful transmission for IA and TDMA+SM as a function of the SINR threshold for some values of the average distance between the cluster centers $D_c$ with perfect CSI.} 
\label{fig:IAandSM_perfectCSI_theory}
\end{figure}

\indent \textbf{Probability of successful transmission with perfect CSI} Assume perfect channel state information and $\gamma_{\rm{o}} = 30$ dB. $\rm{P_s^{IA}}$ and $\rm{P_s^{SM}}$ given by \eqref{eq:ProbSuccessIA_TheoEq} and \eqref{eq:Psuccess_SM_ZF} as a function of the SINR threshold for three values of the average distance between the cluster centers in shown in Fig. \ref{fig:IAandSM_perfectCSI_theory}. As can be seen, for very dense networks, SM outperforms IA which can be explained by the reduced inter-cluster interference due to only a single transmit/receiver pair being active at each cluster. In addition, for moderately dense networks, IA outperforms SM only at high $\theta$ which highlights the nonlinear nature of such comparisons. For very low node density, IA has better performance than SM when the increased distance between transmit/receive pairs does not enforce a zero probability of successful transmission.

\begin{figure}
\centering
\includegraphics[width=4.3in]{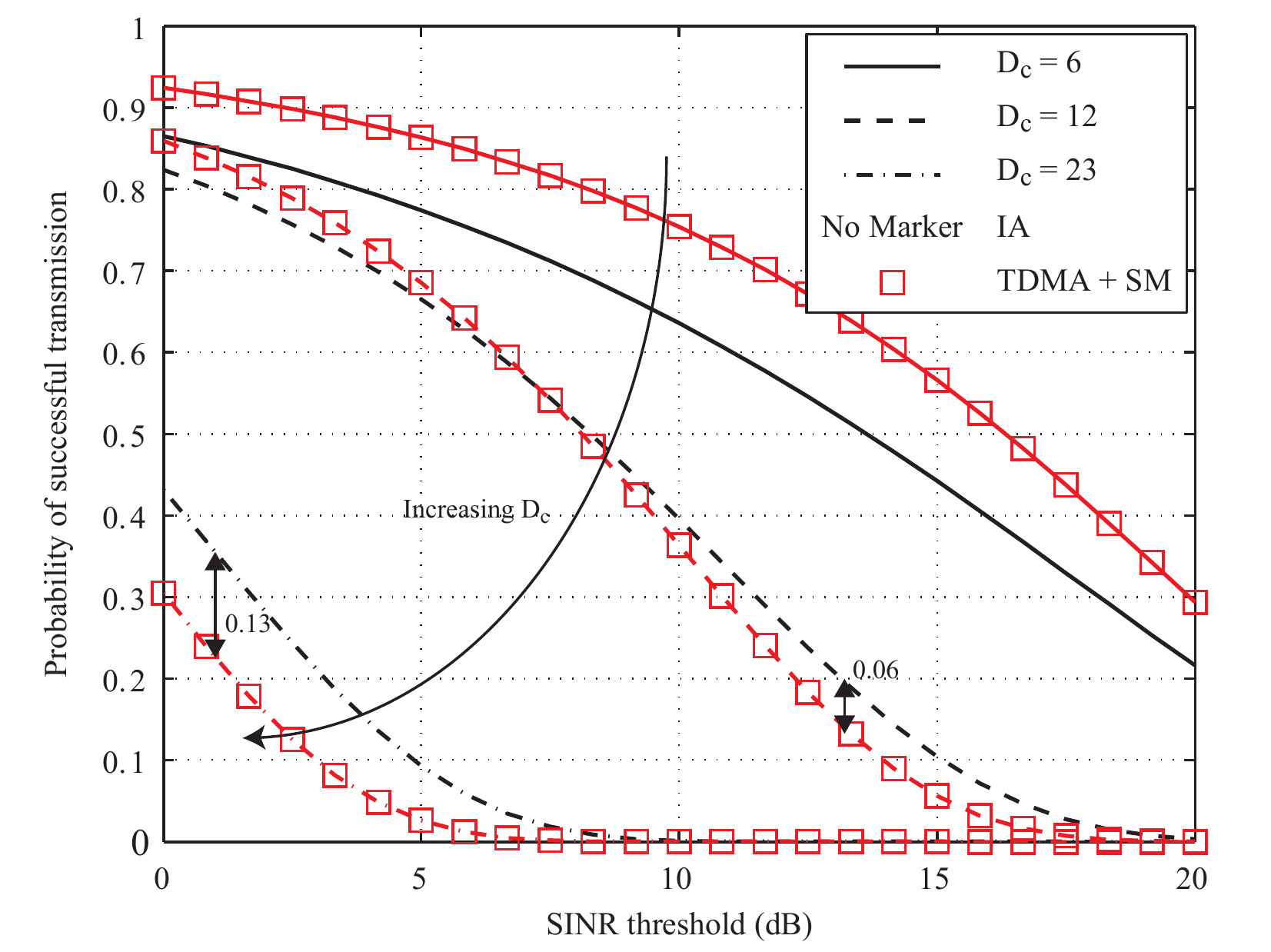}
\caption{$\rm{P_s^{IA}}$ and $\rm{P_s^{SM}}$ as a function of the SINR threshold for some values of the average distance between the cluster centers $D_c$ with the smallest training period of $T_{\rm{t}}=6$.} 
\label{fig:IAandSM_worstCSI_theory}
\end{figure}

\indent \textbf{Probability of successful transmission with imperfect CSI} Now assume the worst channel estimation error variance for IA with the same $\gamma_{\rm{o}} = 30$ dB. The questions of interest are  i) how much the performance of IA will be affected by introducing the imperfect CSI and ii) how does the relative performance between IA and SM change in this scenario. For the same values of $D_c$ and $\theta$ as in Fig. \ref{fig:IAandSM_perfectCSI_theory}, the probabilities of successful transmission for IA and SM when the training period is lowered to $KN=6$ is shown in Fig. \ref{fig:IAandSM_worstCSI_theory}. $\rm{P_s^{IA}}$ and $\rm{P_s^{SM}}$ remain relatively unchanged for dense networks but are reduced for moderate and high node densities. Compared to the perfect CSI case, SM now outperforms IA for a larger range of $\theta$ for moderate node densities. Also, IA has lost most of its advantages at moderate node densities (from maximum of $0.1$ to $0.06$) and the gap between IA and SM has decreased for the widely dispersed network. 
As a reference, the error of \eqref{eq:ProbSuccessIA_TheoEq} when $T_{\rm{t}}=6$ compared with numerical results for the curves shown in Fig. \ref{fig:IAandSM_worstCSI_theory}, in the worst case, is less than $0.0035$.

\begin{figure}
\centering
\includegraphics[width=4.3in]{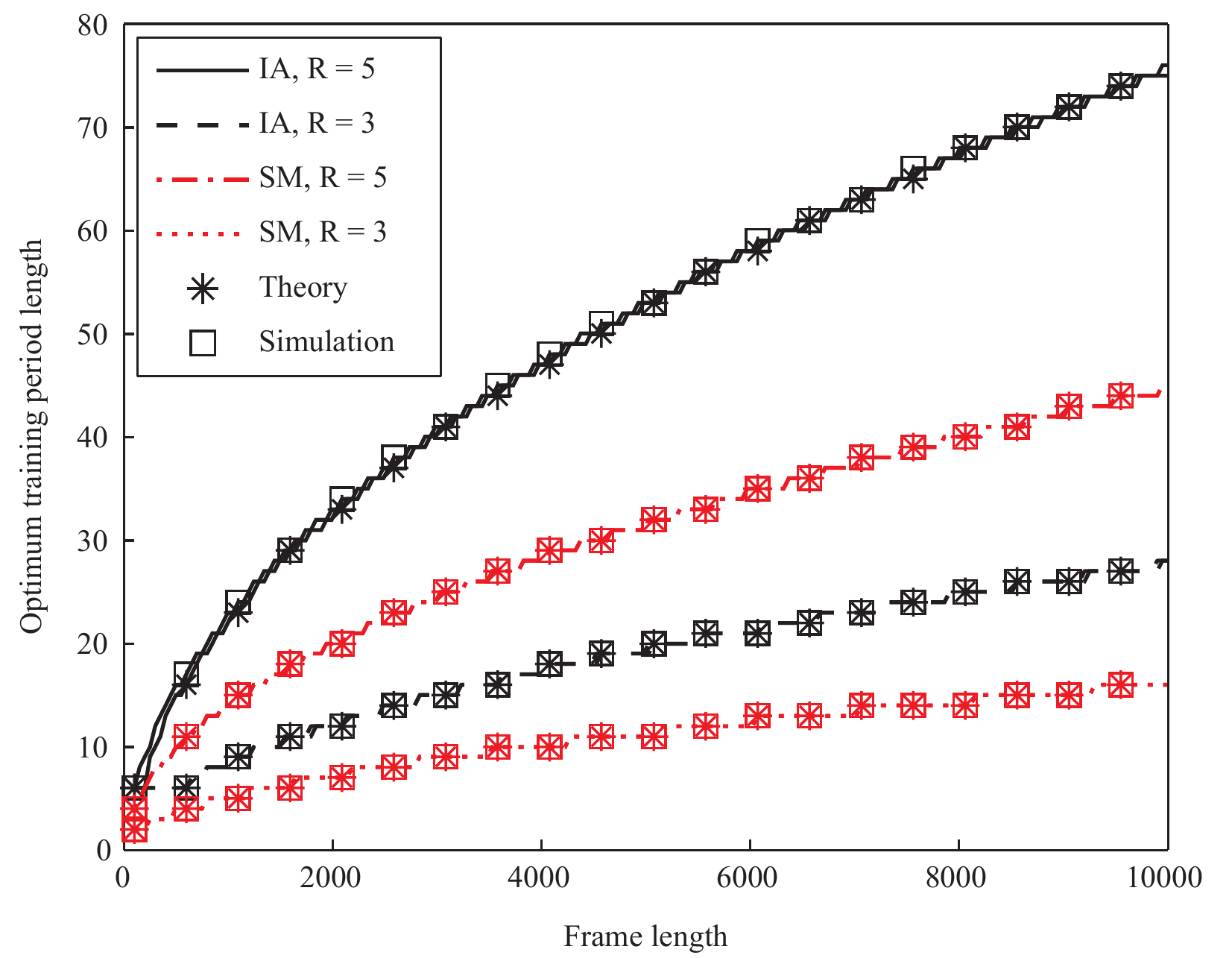}
\caption{The optimum training period as obtained using \eqref{eq:delta_hat_1} and \eqref{eq:delta_hat_SM} for IA and SM together with the corresponding optimum values found through numerically optimizing \eqref{eq:trainingOptimization} and \eqref{eq:trainingOptimization_SM} at two values of cluster radius (and hence average cluster center distance) for $\theta=20$ dB and  $\gamma_{\rm{o}} = 30$ dB as a function of total frame length, $T$.} 
\label{fig:OptimumTraining}
\end{figure}

\indent \textbf{Optimum training period} Again, assume $\gamma_{\rm{o}} = 30$ dB. The optimum training period as obtained using \eqref{eq:delta_hat_1} and \eqref{eq:delta_hat_SM} for IA and SM together with the corresponding optimum values found through numerically optimizing \eqref{eq:trainingOptimization} and \eqref{eq:trainingOptimization_SM} for $\theta=20$ dB and two values of cluster radius (and hence average cluster center distance) as a function of total frame length, $T$, is shown in Fig. \ref{fig:OptimumTraining}. 
As can be seen, \eqref{eq:delta_hat_1} and \eqref{eq:delta_hat_SM} can be used to accurately find the optimum training period for a wide range of node density and total frame length. As expected, IA requires more training than SM but surprisingly, the ratio of the optimum training periods between the two transmission techniques seems to be a constant value independent of the total block length, $T$, which requires further analysis not in the scope of this work.
Also, as the Taylor expansions used to derive \eqref{fig:OptimumTraining} and \eqref{eq:delta_hat_1} are at $T\! =\! \infty$,  small inaccuracies at small frame lengths is expected. Moreover, the numerical errors in numerically calculating the integrals of the coefficients  in \eqref{eq:delta_hat_1} become harder to confine as $R$ (and hence the integration ranges) increases which explains the deviation of the results obtained using \eqref{eq:delta_hat_1} from the true optimum points for $R\! =\!5$.

\begin{figure}
\centering
\includegraphics[width=4.1in]{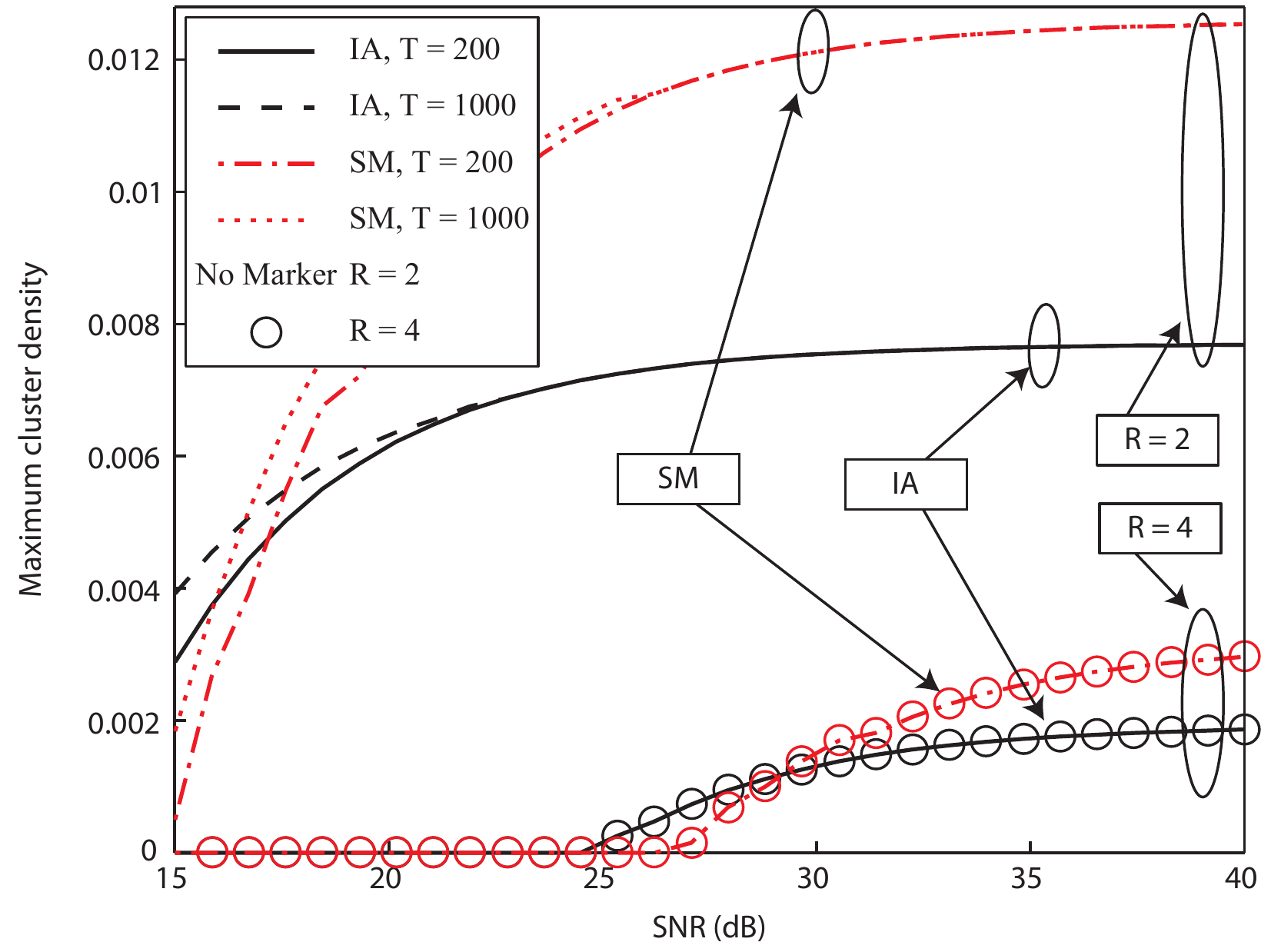}
\caption{The maximum cluster density, $\lambda_P^{\epsilon}$, as given by \eqref{eq:Q_inverse_epsilon_Ns_1_IA} and \eqref{eq:Q_inverse_SM} for two values of cluster radius, $R$, and two values of the total block length, $T$, at the SINR threshold of $\theta=17$ dB and the maximum tolerable outage probability of $\epsilon =0.1$ as a function of transmit SNR, $\gamma_{\rm{o}}$.} 
\label{fig:MaximumClusterDesnity}
\end{figure}

\begin{figure}
\centering
\includegraphics[width=4.3in]{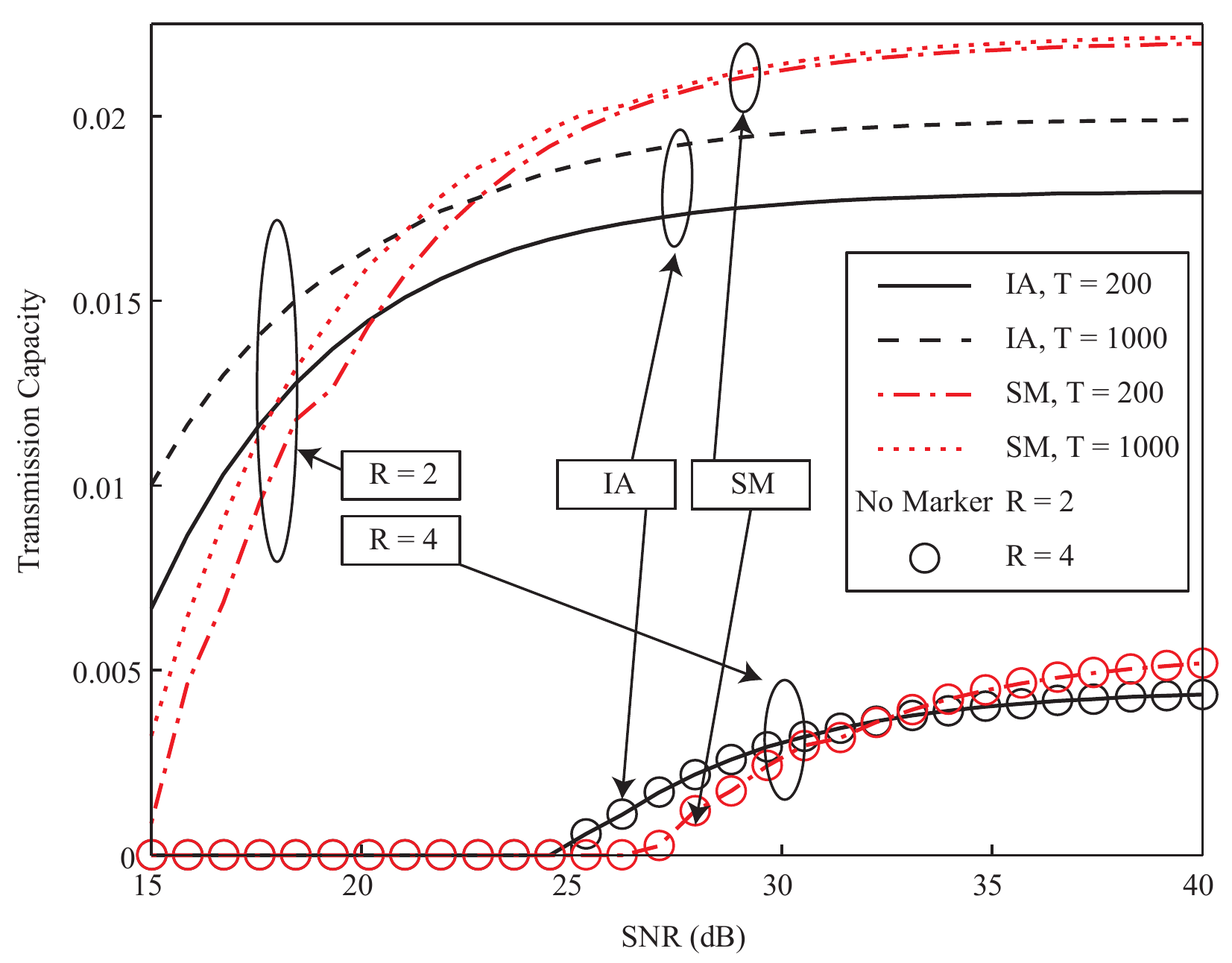}
\caption{The transmission capacity, $c(\epsilon)$, as given by \eqref{eq:TransCap_1} and \eqref{eq:TC_SM} for two values of cluster radius, $R$, and two values of the total block length, $T$, at the SINR threshold of $\theta=17$ dB and the maximum tolerable outage probability of $\epsilon=0.1$ as a function of transmit SNR, $\gamma_{\rm{o}}$. In each case, the optimum training periods as given by \eqref{eq:delta_hat_1} and \eqref{eq:delta_hat_SM} were used to compute the transmission capacities.} 
\label{fig:NetworkTput}
\end{figure}

\indent \textbf{Transmission capacity} Now fix the SINR threshold in \eqref{eq:Q_inverse_epsilon_Ns_1_IA} and \eqref{eq:Q_inverse_SM} to $\theta=17$ dB and fix the maximum tolerable outage probability to $\epsilon=0.1$. The maximum cluster density, $\lambda_P^{\epsilon}$, as given by \eqref{eq:Q_inverse_epsilon_Ns_1_IA} and \eqref{eq:Q_inverse_SM} and the corresponding transmission capacities given by \eqref{eq:TransCap_1} and \eqref{eq:TC_SM} for two values of cluster radius, $R$, and two values of the total block length, $T$, as a function of transmit SNR, $\gamma_{\rm{o}}$, are shown in Fig. \ref{fig:MaximumClusterDesnity} and Fig. \ref{fig:NetworkTput}. 
Note that for $R=5$ only $T=200$ curves are shown as $T=1000$ curves were not conveying any new observations. Also, for both IA and SM, optimum training periods are used to compute the transmission capacity. As seen in Fig. \ref{fig:MaximumClusterDesnity}, except for the high SNR values, increasing $T$ allows for a better channel estimate at the receivers and consequently each receiver can tolerate more inter-cluster interference and hence a large cluster density. Also as the radius of each cluster increases, the channel estimation errors increase and the maximum tolerable cluster density decreases. Although by increasing $T$, the crossing point between IA and SM for  $\lambda_P^{\epsilon}$ shifts to a lower SNR value, the corresponding crossing point for the transmission capacity shifts to higher SNR values which highlights the tradeoff between a higher obtainable throughput by increasing $K$ in each cluster and the accompanying higher accumulated non-aligned interference at each receiver due to the imperfect CSI estimation.

\section{Conclusion} \label{sec:Conclusion}
In this paper, we obtained an expression for the probability of successful transmission in a clustered MIMO IA network when the communication takes place over a Rayleigh fading quasi-static block-fading channel of a finite length and the impact of channel estimation on the accuracy of the obtained CSI is taken into account. We formulated an optimization of the training period length and explicitly solved it for the case of a single stream from each transmitter. We then presented the exact transmission capacity for the same case of a single stream from each transmitter and also provided an upper bound for the general case. Through simulations, we showed that probability of successful transmission and transmission capacity of TDMA+SM can be higher than IA in dense networks or when the mobility of the nodes is high.

\indent Our analyses in this paper were based on a generic IA filter design approach where the only goal of the transmitters is to align the interference without optimizing the direct links. Any such variations of the IA algorithm can potentially increase the gains of IA but quantifying such transmission techniques are out of the scope of this work. Moreover, we assumed a perfect analog feedback link for IA where only the minimum required channel resources are used to relay the exact estimated channel values at the receivers to the rest of the nodes in each cluster. By using the results of \cite{Ayach2012}, it is possible to include the analog feedback analysis but doing so would result in more involved expressions not benefiting the current scope of this work. Other optimization problems, such as optimizing the SINR threshold for the maximum tolerable outage probability, can be readily defined using the results of this paper.

\appendices
\section{Proof of Theorem \ref{thm:one}}\label{AppendixProof:PoutTheorem}
Since $h_{\hat{o}o}\!\sim\!\Gamma(N_s,1)$,  its complementary cumulative distribution function is
$
F(x)= \frac{\Gamma(N_s,x)}{\Gamma(N_s)}=e^{-x}\sum_{k=0}^{N_s-1}\frac{x^k}{k!}.
$
Hence,
\begin{align}
{\rm{P_s^{IA}}}(\theta) &= \PP\left(\frac{g_{\hat{o}o}(1-\beta^2_{\hat{o}o})h_{\hat{o}o}}
{\frac{N_s}{\gamma_{\rm{o}}} + 
I_s +  
I_e +I_i}>\theta\right)
= \PP\left(h_{\hat{o}o}
>
\eta \left(\frac{N_s}{\gamma_{\rm{o}}} + I_s + I_e +I_i\right)
\right)
\nonumber \\
&\stackrel{(a)}{=}\sum_{k=0}^{N_s-1}\frac{\eta^k}{k!}\EE\negmedspace\left[\left( \frac{N_s}{\gamma_{\rm{o}}} +I_{s} +I_{e} +  I_{i}\right)^{k}\negthickspace\! e^{- \eta\left(\frac{N_s}{\gamma_{\rm{o}}} +I_{s} +I_{e} +I_{i}\right)}\right]
\label{eq:Clustering_PsIA_temp0} \\
&\stackrel{(b)}{=}\sum_{k=0}^{N_s-1} \frac{(-\eta)^k}{k!} \frac{\d^k}{\d s^k}\EP e^{-s\left( \frac{N_s}{\gamma_{\rm{o}}}+I_{s} +I_e +I_i\right)}\Big{|}_{s=\eta},
\label{eq:Clustering_PsIA_temp1} 
\intertext{where $(a)$ follows from distribution of $h_{\hat{o}o}$ and $(b)$ from the derivative property of the Laplace transform. Since $I_s$, $I_e$, and $I_i$ in \eqref{eq:Clustering_SINR_temp1} are all independent, \eqref{eq:Clustering_PsIA_temp1} equals} 
{\rm{P_s^{IA}}}(\theta) &= \sum_{k=0}^{N_s-1} \frac{(-\eta)^k}{k!} \frac{\d^k}{\d s^k}
 e^{-s \frac{N_s}{\gamma_{\rm{o}}}}\E e^{-sI_s} \EP e^{-sI_e} \EP e^{-sI_i}\Big{|}_{s=\eta}
\nonumber 
\\
&=\sum_{k=0}^{N_s-1} \frac{(-\eta)^k}{k!} \frac{\d^k}{\d s^k}e^{- s\frac{N_s}{\gamma_{\rm{o}}}} \E e^{-s g_{\hat{o}o}\beta^2_{\hat{o}o} h_{\hat{o}o} }\L^{!o}_{\I_{e}}(s)\L_{\I_{i}}(s)\Big{|}_{s=\eta}, \label{eq:Pout_IA_Original}
\end{align}
where $\L^{!o}_{\I_e}(s)$ is the Laplace transform of the intra-cluster interference w.r.t the reduced Palm measure and $\L_{\I_i}(s)$  is the Laplace transform of the inter-cluster interference. Since $h_{\hat{o}o} \sim \Gamma(N_s,1)$, using \eqref{eq:Clustering_EstimationVarianceAndSNR},
\begin{align}
\E e^{-s g_{\hat{o}o}\beta^2_{\hat{o}o} h_{\hat{o}o} }= \left(\frac{\frac{T_{\rm{t}}\gamma_{\rm{o}}}{N}+D_{\rm{r}}^\alpha}{s+\frac{T_{\rm{t}}\gamma_{\rm{o}}}{N}+D_{\rm{r}}^\alpha}\right)^{N_s}. \label{eq:coeff_for_imperfectCSI}
\end{align}
We now evaluate $\L^{!o}_{\I_e}(s)$
\begin{align}
\L^{!o}_{\I_{e}}(s)&= \EP \left[e^{-s\sum\limits_{z\in \boldsymbol{\Psi}_o}g_{\hat{o}z}\beta^2_{\hat{o}z} h_{\hat{o}z} }\right] =\EP \left[\prod_{z\in \boldsymbol{\Psi}_o}e^{-s g_{\hat{o}z}\beta^2_{\hat{o}z} h_{\hat{o}z} }\right]
\nonumber \\
&\stackrel{(a)}{=} \EP \left[\prod_{z\in \boldsymbol{\Psi}_o}\left(\frac{1}{1+s g_{\hat{o}z}\beta^2_{\hat{o}z} }\right)^{N_s}\right]
\nonumber \\
&=\EP
\left[\prod_{z\in \boldsymbol{\Psi}_o}\left(\frac{\frac{T_{\rm{t}}\gamma_{\rm{o}}}{N}+\|z-\hat{o}\|^\alpha}
{s+\frac{T_{\rm{t}}\gamma_{\rm{o}}}{N}+\|z-\hat{o}\|^\alpha}\right)^{ N_s}\right], \label{eq:tempProof_4}
\end{align}
where $(a)$ follows from the Laplace transform of $h_{\hat{o}z}$ and in the last step $\|x\|^{-\alpha}$ was substituted for the path loss.  Observe that \eqref{eq:tempProof_4} is the probability generating functional (PGF) of the representative cluster $\boldsymbol{\Psi}_o$ w.r.t its reduced Palm probability. We now use the following  result from \cite[Lemma 1]{Ganti2009} which we state for completeness. For any function $f(x):\R^2\to \R^+$,
\begin{align}
\EP\prod_{x\in \boldsymbol{\Psi}_o}f(x) =\frac{1}{\pi R^2}\int_{A=B(o,R)}\left(\frac{1}{\pi R^2}\int_{A=B(o,R)}f(x-y)\d x\right)^{K-1}\d y.
\label{eq:Temp_Lie_1}
\end{align}
Using \eqref{eq:Temp_Lie_1}, \eqref{eq:tempProof_4} equals
\begin{align}
\L^{!o}_{\I_e}(s)&= 
\frac{1}{\pi R^2}\int\limits_{\!\!B(o,R)}
\negmedspace
\Biggg[\!\frac{1}{\pi\! R^2}
\negthickspace\!
\int\limits_{\!\!B(o,R)} \negthickspace\negthickspace\negmedspace\left(\!\frac{\frac{T_{\rm{t}}\gamma_{\rm{o}}}{N}\!+\!\|x\!-\!y\!-\!\hat{o}\|^\alpha}{s\!+\!\frac{T_{\rm{t}}\gamma_{\rm{o}}}{N}\!+\!\|x\!-\!y\!-\!\hat{o}\|^\alpha}\!\right)^{\!\!\!N_s}
\negmedspace\negmedspace
\d x\Biggg]^{\!K\!-\!1}
\negthickspace\negthickspace\negthickspace\!
\d y.
\label{eq:lap1_temp}
\intertext{We now focus on $\L_{\I_i}(s)$}
\L_{\I_{i}}(s)&= \EP \left[e^{-s\sum_{x\in \boldsymbol{\Phi}}g_{\hat{o}x} h_{\hat{o}x} }\right]
= \EP \left[\prod_{x\in \boldsymbol{\Phi}}\left(\frac{1}{1+s\|x-\hat{o}\|^{-\alpha}}\right)^{N_s}\right] \nonumber 
\\
&\stackrel{(a)}{=}\!\exp\!\Bigg(
\negthickspace\negthickspace
-\lambda_p \! \int\limits_{\R^2}
\negthickspace\!
1\negmedspace-\!\negmedspace\Bigg[\!\frac{1}{\pi\! R^2}
\negthickspace\!\!
\int\limits_{B(o,R)}
\negthickspace\negthickspace\negmedspace\!
\left(\!\frac{\|x\!-\!y\|^{\alpha}}{s\!+\!\|x\!-\!y\|^{\alpha}}\!\!\right)^{\negthickspace\! N_s}
\negthickspace\negthickspace\negmedspace
\d x\Bigg]^{\!K}
\negthickspace
\d y
\Bigg), \label{eq:lap2_temp}
\end{align}
where $(a)$ follows from \eqref{eq:ProbGenFunctional_1}  which characterizes the PGF of a Poisson cluster process\cite{Stoyan2008}
\begin{align}
\E \! \prod_{x\in \boldsymbol{\Phi}} \! f(x) 
\!=\! \exp
\!\Biggg(\negthickspace\!
-\lambda_p\int\limits_{\R^2}1-\Bigg(\frac{1}{\pi R^2}
\negthickspace\negthickspace
\int\limits_{B(o,R)}
\negthickspace\negthickspace\!
f(x-y)\d x\Bigg)^{\!\!K} \!\d y\Biggg).
\label{eq:ProbGenFunctional_1}
\end{align}
Note that in \eqref{eq:lap1_temp}, $x$ and $y$ vary within a cluster, while in \eqref{eq:lap2_temp}, $y$ varies over the whole plane. Also, in \eqref{eq:lap1_temp}, the interference is only from the other $K\!-\!1$ transmitters in the cluster and hence the exponent of $K\!-1$, while in \eqref{eq:lap2_temp}, the interference is from all the $K$ transmitters at each cluster and hence the exponent of $K$.   

\section{Proof of Lemma \ref{lem:UpperBoundForThePoutTerms}} \label{AppendixProof:UpperBoundForThePoutTerms}
Define the following PDF
\begin{align}
f_p(x) =
\left\{
\begin{array}{ll}
\frac{1}{\pi R^2} &  x \in B(o,R) \\
0 & \text{Otherwise}
\end{array}\right. .
\label{eq:simplifying_LIi_temp0}
\end{align}
Now, using \eqref{eq:simplifying_LIi_temp0}
\begin{align}
\L_{\I_i}(s)&= \exp\left(-\lambda_p \int_{\R^2}1-\left[\frac{1}{\pi R^2}\int_{B(o,R) } \E_h[ e^{-sh\|x-y\|^{-\alpha}}]\d x\right]^{K}\d y\right) \nonumber \\
&= \exp\left(-\lambda_p \int_{\R^2}1-\left[\int_{\R^2 } \E_h[ e^{-sh\|x-y\|^{-\alpha}}]f_p(x)\d x\right]^{K}\d y\right).
\label{eq:simplifying_LIi_temp1}
\end{align}
Therefore, the inner integral in \eqref{eq:simplifying_LIi_temp1} can be seen as the expectation of $\E_h[ e^{-sh\|x-y\|^{-\alpha}}]$ w.r.t $f_p(x)$. Using Jensen's inequality, as $\left(\E_x [z] \right)^K \leq \E_x [z^K]$ for $K>2$, it follows from \eqref{eq:simplifying_LIi_temp1} that
\begin{align}
\L_{\I_i}(s) &\leq \exp\left(-\lambda_p \int_{\R^2}1-\int_{\R^2 } f_p(x)\E_h[ e^{-sh\|x-y\|^{-\alpha}}]^{K}\d x\d y\right)
\nonumber \\
& \stackrel{(a)}{=} \exp\left(-\lambda_p \int_{\R^2}\int_{\R^2 } \left(1- \left(\frac{ \|x-y\|^{\alpha}}{s + \|x-y\|^{\alpha}}\right)^{KN_s}\right)f_p(x)\d x\d y\right)
\nonumber \\ 
& = \exp\left(-\lambda_p \int_{\R^2}f_p(x)\int_{\R^2 } \left(1- \left(\frac{ \|y^\prime\|^{\alpha}}{s + \|y^\prime\|^{\alpha}}\right)^{KN_s}\right)\d y^\prime\d x\right)
\nonumber \\
& = \exp\left(-\lambda_p s^{2/\alpha}\frac{\Gamma(KN_s+2/\alpha)\Gamma(1-2/\alpha)}{\Gamma(KN_s)} \int_{\R^2}f_p(x)\d x\right)
\nonumber \\
& = \exp\left(-\lambda_p s^{2/\alpha}\frac{\Gamma(KN_s+2/\alpha)\Gamma(1-2/\alpha)}{\Gamma(KN_s)} \right), \nonumber
\end{align}
where $(a)$ follows from $h\sim \Gamma(N_s,1)$. Similarly for $\L^{!o}_{\I_e}$
\begin{align}
\L^{!o}_{\I_e}(s) & = \frac{1}{\pi R^2}
\negthickspace\negthickspace
\int\limits_{B(o,R)}
\left[\frac{1}{\pi R^2}
\negthickspace\negthickspace
\int\limits_{B(o,R) } \E_h[e^{-shf(\|x-y-\hat{o}\|)}]\d x\right]^{K-1}
\negthickspace\negthickspace\negthickspace\negthickspace
\d y 
\nonumber \\
 &\stackrel{(a)}{\leq} \E_{x,y}\left[\left(\E_h[e^{-shf(\|x-y-\hat{o}\|)}]\right)^{K-1}\right],
\label{eq:temp4}
\intertext{where $f(x)= \frac{1}{\|x\|^\alpha +\frac{T_{\rm{t}}\gamma_{\rm{o}}}{N}}$, $x$ and $y$ are two points independently and uniformly distributed in $B(o,R)$, and $(a)$ follows from the Jensen's inequality. Evaluating the expectation w.r.t $h$, \eqref{eq:temp4} equals}
\L^{!o}_{\I_e}(s) &\leq \E\left[ \frac{1}{(1+sf(\|x-y-\hat{o}\|))^{N_s(K-1)}}\right]. \label{eq:temp_bound1}
\end{align}
Let $(r,\theta)$ be the polar representation of $x-y$, where $x$ and $y$ are two points independently and uniformly distributed on $B(o,r)$. Then $\theta \sim U(0,2\pi)$ and from \cite{hammersley1950distribution} the PDF of $r$ is given by
\eqref{eq:PDFofR_Uniform}. Note that \eqref{eq:PDFofR_Uniform} equals
$I_{1-\frac{x^2}{4R^2}}(3/2,1/2) = \frac{2}{\pi}\left(\arcsin(\sqrt{1-x^2/4R^2})-\frac{x}{2R}\sqrt{1-x^2/4R^2}\right)$.
Without loss of generality, we can assume that $\hat{o} = (D_{\rm{r}},0)$. Then \eqref{eq:temp_bound1} simplifies to 
\eqref{eq:L_Ie_UpperBound}.

\section{Proof of Lemma \ref{lem:UpperBoundOnPs_noDerivitive}} \label{AppendixProof:UpperBoundOnPs_noDerivitive}
From \eqref{eq:Clustering_PsIA_temp0} we have
\begin{align}
{\rm{P_s^{IA}}} &= \sum_{k=0}^{N_s-1}\frac{\eta^k}{k!}\EE\negmedspace\left[\left( \frac{N_s}{\gamma_{\rm{o}}} +I_{s} +I_{e} +  I_{i}\right)^{k}e^{- \eta\left(\frac{N_s}{\gamma_{\rm{o}}} +I_{s} +I_{e} +I_{i}\right)}\right]
\nonumber \\
&= \sum_{k=0}^{N_s-1}\frac{\eta^k}{k!}\EE\negmedspace\left[\left( \frac{N_s}{\gamma_{\rm{o}}} +I_{s} +I_{e} +  I_{i}\right)^{k} 
e^{-\left(\frac{k}{e} + \eta-\frac{k}{e}\right)\left(\frac{N_s}{\gamma_{\rm{o}}} +I_{s} +I_{e} +I_{i}\right)}\right]
\nonumber \\
&\stackrel{(a)}{\leq} \sum_{k=0}^{N_s-1}\frac{\eta^k}{k!}\EE\negmedspace\left[
e^{- \left(\eta-\frac{k}{e}\right)\left(\frac{N_s}{\gamma_{\rm{o}}} +I_{s} +I_{e} +I_{i}\right)}\right]
\nonumber \\
&=\sum_{k=0}^{N_s-1}\frac{\eta^k}{k!}e^{-(\eta-k/e)\frac{N_s}{\gamma_{\rm{o}}}}
\E e^{-(\eta - k/e)I_{s}}
\EE e^{-(\eta - k/e)I_{e}}
\E e^{-(\eta - k/e)I_{i}},
\label{eq:proof_lemma1_temp1}
\end{align}
where $(a)$ follows from the observation $x^ke^{-\frac{k}{e}x}\leq 1$ for $x>0$
and the proof follows by comparing  \eqref{eq:proof_lemma1_temp1} to \eqref{eq:Pout_IA_Original} and making the necessary change of variables.

\section{Coefficients of \eqref{eq:TrainingOpt_Ns1_fdFormulation_3}} \label{Appendix:C1_and_C2}
\vspace{-0.3in}
\begin{align}
C_1 
&= \frac{\theta D_{\rm{r}}^\alpha f_d N}{2 \gamma_{\rm{o}}^4} \Big(-2 \theta D_{\rm{r}}^{2 \alpha} f_d K N \gamma_{\rm{o}} \left(A_4 K \lambda_p \gamma_{\rm{o}}+1\right) 
\nonumber \\
&+2 \gamma_{\rm{o}}^2 \left(-A_1 f_d (K-1) N + K\gamma_{\rm{o} }\left(f_d K^2 N-1\right) \right)
\nonumber \\
&-D_{\rm{r}}^\alpha \gamma_{\rm{o}}^2 \left(2+2 A_4 K \lambda_p \gamma_{\rm{o}}+f_d N \left(2+ K (K+1)\left(\theta-2\right) -2 A_4 K^3 \lambda_p \gamma_{\rm{o}}\right)\right)
\nonumber \\
&-\theta D_{\rm{r}}^{3 \alpha} f_d N (1+K \lambda_p \gamma_{\rm{o}} (2 A_4+(A_5+A_4 (2+A_4 K \lambda_p)) \gamma_{\rm{o}}))\Big), \nonumber
\\
C_2 &= 
\frac{N^2 \theta {D_{r}}^\alpha {f_{d}}^2}{2 \gamma_{\rm{o}}^4} \Big(2 \gamma_{\rm{o}}^2 {D_{r}}^\alpha + \theta {D_{r}}^{3 \alpha} - 2 A_{1} \gamma_{\rm{o}}^2 - 2 K \gamma_{\rm{o}}^2 {D_{r}}^\alpha + 2 A_{1} K \gamma_{\rm{o}}^2 
\nonumber \\
&+ K^2 \gamma_{\rm{o}}^2 \theta {D_{r}}^\alpha + 2 K \gamma_{\rm{o}} \theta {D_{r}}^{2 \alpha} + K \gamma_{\rm{o}}^2 \theta {D_{r}}^\alpha 
\nonumber \\
& + K\lambda_p \theta {D_{r}}^{3 \alpha} \left({A_{4}}^2 K \gamma_{\rm{o}}^2 {\lambda_{p}} + 2 A_{4} K \gamma_{\rm{o}}^2  {D_{r}}^{- \alpha}  + 2 A_{4}  \gamma_{\rm{o}}  + 2 A_{3}  \gamma_{\rm{o}}^2 + A_{5} \gamma_{\rm{o}}^2 \right)\Big), \nonumber
\intertext{where}
A_1 &=  \frac{1}{\pi R^2}\negthickspace\negthickspace\int\limits_{\!\!B(o,R)}
\negthickspace\negmedspace
\frac{1}{\pi\! R^2}
\negthickspace\negthickspace\!
\int\limits_{\!\!B(o,R)} \negthickspace\negmedspace
\|x\!-\!y\!-\!\hat{o}\|^\alpha
\d x \d y, \nonumber
\quad 
A_2 = \|x\!-\!y\|^\alpha + \theta D_{\rm{r}}^{\alpha}, \nonumber
\\
A_3 &= \int\limits_{\mathbb{R}^2}
\negthickspace\Biggg[
\frac{1}{\pi\! R^2}
\negthickspace\negthickspace\!
\int\limits_{\!\!B(o,R)} \negthickspace\negmedspace
\frac{\|x\!-\!y\|^\alpha}{\left(A_2\right)^3}
\d x \Biggg]
\Biggg[
\frac{1}{\pi\! R^2}
\negthickspace\negthickspace\!
\int\limits_{\!\!B(o,R)} \negthickspace\negmedspace
\frac{\|x\!-\!y\|^\alpha}{A_2}
\d x
\Biggg]^{K-1}
\negthickspace\negthickspace\negthickspace\negthickspace \d y, \nonumber
\\
A_4 &= \int\limits_{\mathbb{R}^2}
\negthickspace\Biggg[
\frac{1}{\pi\! R^2}
\negthickspace\negthickspace\!
\int\limits_{\!\!B(o,R)} \negthickspace\negmedspace
\frac{\|x\!-\!y\|^\alpha}{\left(A_2\right)^2}
\d x \Biggg]
\Biggg[
\frac{1}{\pi\! R^2}
\negthickspace\negthickspace\!
\int\limits_{\!\!B(o,R)} \negthickspace\negmedspace
\frac{\|x\!-\!y\|^\alpha}{A_2}
\d x
\Biggg]^{K-1}
\negthickspace\negthickspace\negthickspace\negthickspace \d y, \nonumber
\\
A_5 &= (K-1)\int\limits_{\mathbb{R}^2}
\negthickspace\Biggg[
\frac{1}{\pi\! R^2}
\negthickspace\negthickspace\!
\int\limits_{\!\!B(o,R)} \negthickspace\negmedspace
\frac{\|x\!-\!y\|^\alpha}{\left(A_2\right)^2}
\d x \Biggg]^2
\Biggg[
\frac{1}{\pi\! R^2}
\negthickspace\negthickspace\!
\int\limits_{\!\!B(o,R)} \negthickspace\negmedspace
\frac{\|x\!-\!y\|^\alpha}{A_2}
\d x
\Biggg]^{K-2}
\negthickspace\negthickspace\negthickspace\negthickspace \d y. \nonumber
\end{align}

\section{Coefficients of \eqref{eq:TrainingOpt_Ns1_fdFormulation_simple_1}} \label{Appendix:D1_and_D2}
\vspace{-0.3in}
\begin{align}
D_1 & = \frac{f_d\theta D_{\rm{r}}^{\alpha} }{\frac{\gamma_{\rm{o}}}{N}}\left(
-K \!+\! 
\frac{f_d}{\frac{\gamma_{\rm{o}}}{N}}\left(
  \frac{K^3N \gamma_{\rm{o}}}{N}
\!-\! D_{\rm{r}} ^{\alpha}(\theta\!+\! 1)
\!-\!\frac{\theta D_{\rm{r}} ^{\alpha}}{2}(K\!-\!1)(K\!+\!2)
\!-\!(K\!-\!1) A_1
\right)
\right), \nonumber \\
D_2 & = 
\frac{f_d^2\theta D_{\rm{r}} ^{\alpha}}{\left(\frac{\gamma_{\rm{o}}}{N}\right)^2}\left(
 D_{\rm{r}} ^{\alpha}(\theta +1) 
+\frac{\theta D_{\rm{r}} ^{\alpha}}{2}(K-1)(K+2)
+(K-1) A_1
\right), \nonumber
\end{align}
where $A_1$ is given in Appendix \ref{Appendix:C1_and_C2}.

\section{Coefficients of \eqref{eq:TrainingOpt_SM_fdFormulation_1}} \label{Appendix:B1_and_B2}
\vspace{-0.3in}
\begin{align}
B_1 &=  -B_2
- \frac{ D_{r}^\alpha f_{d}}{\gamma_{\rm{o}}^2 \alpha} \left(N \alpha \theta D_{r}^\alpha + N \gamma_{\rm{o}} \alpha \theta + 2 \mathcal{J} \gamma_{\rm{o}} \theta^{\frac{2}{\alpha}} D_{r}^2 \lambda_p\right), \nonumber
\\
B_2 
&= \frac{D_{r}^{2\alpha}f_{d}^2}{2\gamma_{\rm{o}}^4 \alpha^2} \left({\left(
\alpha N \theta  \left(\gamma_{\rm{o}} + D_{r}^\alpha\right) 
+ 2\gamma_{\rm{o}} \mathcal{J} \theta^{\frac{2}{\alpha}} D_{r}^{ 2} \lambda_p
\right)}^2 
- 2\gamma_{\rm{o}}^2 \alpha \mathcal{J} \theta^{\frac{2}{\alpha}} D_{r}^{ 2} \lambda_p \left(2/\alpha - 1\right)
\right), \nonumber
\end{align}
where $\mathcal{J}$ is defined in \eqref{eq:Psuccess_SM_ZF}.

\bibliographystyle{IEEEtran}
\bibliography{IEEEabrv,../../Bib/Thesis}
\newpage

\begin{table}[!tb]
\begin{center}
\caption{Notation used in this paper}
\label{table:Notation}
\setlength{\extrarowheight}{2pt}
\begin{tabularx}{\linewidth}{|l||X|} 
\hline  
Symbol  & Explanation
\\ \hline \hline
$\mathbf{a}$ \& $\mathbf{A}$
& $\mathbf{a}$ is a vector and $\mathbf{A}$ is a matrix  			\\ \hline
$\mathbf{A}^*$, $\mathbf{A}^T$, \& $\mathbf{A}^{-1}$		
& Conjugate transpose, transpose, and inverse of $\mathbf{A}$		\\ \hline
$\mathbf{A}_{n\times m}$
& $\mathbf{A}$ is a matrix of dimension $n\times m$ 				\\ \hline
$\mathbf{A}(n,m)$
& Element on the $n${th} row and $m${th} column of $\mathbf{A}$	 	\\ \hline
${\rm{rank}}\left(\mathbf{A}\right)$
& Rank of $\mathbf{A}$												\\ \hline
$[ a ]$
& Closest integer to $a$ 											\\ \hline	
$\mathbf{I}_N$ \& $\mathbf{0}_{a\times b}$
& $N\times N$ identity matrix and $a\times b$ matrix of all zeros	\\ \hline
$B(o,R)$
& The disk of radius $R$ centered on the origin 					\\ \hline
$\PP$
& The reduced Palm probability measure 								\\ \hline
$\mathbf{A}_{i,j}$ vs.\ $\mathbf{A}$
& When $\mathbf{A}_{i,j}$ correspond to $i$ and $j$ nodes, $\mathbf{A}$ is a generic random variable i.i.d.\ to the specific ones																\\ \hline
\end{tabularx}
\end{center}
\end{table}

\begin{table}[!tb]
\begin{center}
\caption{Important symbols defined in this paper}
\label{table:symbols}
\setlength{\extrarowheight}{2pt}
\begin{tabularx}{\linewidth}{|l||X|} 
\hline  
$P_{\rm{A}}$
& Channel access probability of each cluster						\\ \hline
$\boldsymbol{\Phi}_{\rm{p}}$
& Spatial locations of the cluster centers with density $\tilde{\lambda}_{\rm{p}}$ \\ \hline
$K$ 
& Number of transmit/receive pairs in each cluster					\\ \hline
$N$
& Number of antennas at each node									\\ \hline
$N_{s}$
& Number of streams from each transmitter							\\ \hline
$\boldsymbol{\Phi}$
& The spatial locations of the potential transmitters with density 	$K\tilde{\lambda}_{\rm{p}}$		\\ \hline
$\boldsymbol{\Psi}_o$
& The cluster to which the transmitter at the origin belongs by		\\ \hline
$\lambda_{\rm{p}}$
& Density of the active transmitters (equals $P_{\rm{A}}\tilde{\lambda}_{\rm{p}}$)	\\ \hline
$D_{\rm{r}}$
& Distance of a receiver from its corresponding transmitter			\\ \hline
$D_{c}$
& Average distance between the cluster centers		\\ \hline
$R$
& Radius of each cluster							\\ \hline
$\beta_{\hat{x}z}^2$ \& $g_{\hat{x}z}$
& Variance of the channel estimation error and pathloss for the link between transmitter $z$ and receiver $\hat{x}$ \\ \hline
$T$, $f_d$, $T_{{\rm{t}}}$, \& $T_{{\rm{f}}}$
& Length of the constant channel block, Doppler frequency ($f_d\approx \frac{1}{T}$), training period, and feedback period \\ \hline
$\theta$
& SINR threshold for successful transmission 						\\ \hline
\end{tabularx}
\end{center}
\end{table}

\end{document}